\documentclass[%
 reprint,
 amsmath,amssymb,
 aps,
]{revtex4-2}

\usepackage[dvipsnames]{xcolor}

\usepackage{enumitem}
\usepackage{lineno}
\usepackage{graphicx}
\usepackage{dcolumn}
\usepackage{bm}
\usepackage{hyperref}

\begin{document}

\preprint{APS/123-QED}

\title{Data-driven dimensionality reduction and causal inference in spatiotemporal climate fields}

\author{Fabrizio Falasca}
\email{fabri.falasca@nyu.edu}
\author{Pavel Perezhogin}
\author{Laure Zanna}
\affiliation{%
Courant Institute of Mathematical Sciences \\ New York University, New York, NY, USA 
}%

\date{\today}

\begin{abstract}

We propose a data-driven framework to describe spatiotemporal climate variability in terms of a few entities and their causal linkages. Given a high-dimensional climate field, the methodology first reduces its dimensionality into a set of regionally constrained patterns. Causal relations among such patterns are then inferred in the \textit{interventional} sense through the fluctuation-response formalism.  To distinguish between true and spurious responses, we propose a novel analytical null model for the fluctuation-dissipation relation, therefore allowing for uncertainty estimation at a given confidence level. We showcase the methodology on the sea surface temperature field from a state-of-the-art climate model. The usefulness of the proposed framework for spatiotemporal climate data is demonstrated in several ways. First, we focus on the correct identification of known causal relations across tropical basins. Second, we show how the methodology allows us to visualize the cumulative response of the whole system to climate variability in a few selected regions. Finally, each pattern is ranked in terms of its ``causal strength'', quantifying its relative ability to influence the system's dynamics. We argue that the methodology allows us to explore and characterize causal relations in spatiotemporal fields in a rigorous and interpretable way.\\

\textbf{This is a post-peer-review, pre-copyedit version of an article published in Physical Review E. The final authenticated version is
available online at:} \href{https://journals.aps.org/pre/abstract/10.1103/PhysRevE.109.044202}{https://journals.aps.org/pre/abstract/10.1103/PhysRevE.109.044202}.









\end{abstract}

\maketitle


\section{Introduction} \label{sec:intro}

The Earth's climate is a complex dynamical system composed by many components, such as the atmosphere and hydrosphere, and their interactions \cite{Gupta}. Such linkages give rise to nontrivial feedbacks, generating self-sustained spatiotemporal patterns \cite{GhilLucarini,LucariniChekroun}. An example is the El Ni\~no Southern Oscillation (ENSO), a recurrent pattern of natural variability emerging from air-sea interaction in the tropical Pacific Ocean \cite{Philander,ENSOcomplexity}. Other examples include the Asian Monsoon, the Indian Ocean Dipole, and the Atlantic Ni\~no, just to cite a few \cite{Wang,Heydt, Webster}. Such patterns, commonly referred to as \textit{modes of variability}, interact with each other on a vast range of spatial and temporal scales \cite{Wallace,Alexander,Chiang}. 
Spatiotemporal climate dynamics can then be thought of as a collection of modes of variability and their linkages, or as commonly referred to, a ``climate network'' \cite{Tsonis,Donges}. The identification of such a complex array of interactions and the quantification of its response to external forcings (e.g., \cite{Julien1,Falasca}) is a fundamental (but nontrivial) problem at the root of our understanding of climate dynamics. It requires hierarchies of models, theories, observations, and new tools to analyze and simplify the description of high-dimensional, complex data \cite{Bracco, LucariniChekroun}. In fact, the exponential growth of data from models and observations, together with appropriate and rigorous frameworks, promise new ways to explore and ultimately understand climate dynamics \cite{Bracco}. An important step when ``learning'' from climate data is to infer meaningful linkages among time series, whether among modes of variability or other features of the system (e.g., global averages). Traditionally, this has been done by quantifying \textit{pairwise} similarities, whether linear or nonlinear (for example \cite{ManuNPGO, FalascaEPJ, Falasca} and \cite{NetworkBackbone}, respectively). Such statistical similarities cannot quantify what we refer to as ``causality'', limiting our ability to discover meaningful mechanisms in high-dimensional dynamical systems such as climate. In the context of dynamical systems, the main idea of causal inference can be informally summarized as follows: given the time series $x_{1}(t),x_{2}(t),...,x_{N}(t)$ of a $N$-dimensional system $\boldsymbol{x}(t) \in \mathbb{R}^{N}$, where $t$ is a time index, we aim in quantifying (a) to what extent and (b) at what time scales changes in a variable $x_{j}(t)$ can influence another variable $x_{k}(t+\tau)$ at later times \cite{Lucarini2018,Baldovin}.\\

This study proposes a scalable framework to (a) coarse grain a spatiotemporal climate field into a set of a few patterns and (b) infer their causal linkages. Altogether, the proposed strategy allows us to study complex, high-dimensional climate dynamics in an interpretable and simplified way.\\

Causality is a fundamental topic in science ranging from foundational questions in physics and philosophy \cite{Hume,Russell,Bohm,Cartwright,Rovelli,Adlam,Jenann,Ismael,IsmaelNew} to practical design and implementation of inference algorithms \cite{PearlBook}. In the last decades, there has been a great interest in developing new methodologies to infer causal associations directly from data. In the case of time series data, attempts to infer causal connections start from the work of Granger \cite{Granger}, who framed the problem of causal inference in terms of prediction. The main idea of Granger causality was to draw a causal link between two variables $x_{j}$ and $x_{k}$ if the past of $x_{j}$ would enhance the predictability of the future of $x_{k}$. Another attempt, coming from the dynamical system literature, was based on the concept of transfer entropy \cite{Schreiber,JakobEntropy}. Crucially, as noted in \cite{Baldovin}, Granger causality and transfer entropy give similar information and are equivalent for Gaussian variables \cite{Barnett}. In the last decades, new ideas from computer science, mainly driven by Pearl \cite{PearlBook,Pearl1}, have given us practical ways to design and implement causal tools mainly based on graphical models. Frameworks of such kind have been recently developed in climate science with contributions ranging from the work of Ebert-Uphoff and Deng (2012) in \cite{Deng} to the newer ``PCMCI'' method led by Runge et al. (2019) \cite{Runge}; see \cite{Camps} for a review. Additionally, the Machine Learning (ML) community is actively interested in causality and applications and we refer to \cite{Matt} for details on new developments and open problems in ``Causal ML''.\\ 

Recently, it has been noted that linear response theory \cite{KuboFDT,Marconi} serves as a rigorous framework to understand causality in physical systems \cite{Ruelle,LucariniColangeli,Aurell,Lucarini2018,Valerio2021,Baldovin}. The main rationale is that in physics, causal effects can be identified by observing the response of the system to \textit{external} actions \cite{Barnett,Aurell}. In the limit of infinitesimally small external perturbations, the linear response formalism provides a strategy to compute changes in statistical properties of a physical system solely from the notion of the \textit{unperturbed} dynamics \cite{MajdaBook,Valerio2021}. This allows us to capture causal relations from data in the \textit{interventional} sense \cite{Pearl1,Aurell,Baldovin,IsmaelNew}.\\

The fluctuation-response formalism \cite{Baldovin} differs from many commonly employed causal algorithms, such as conditional independence testing \cite{RungeScience}, Granger causality \cite{Granger} and transfer entropy \cite{Schreiber}, by focusing directly on the problem of causal effect estimation \cite{PearlBook,Manshour} rather than causal discovery (i.e., direct causal links) \cite{giorgia}. Many causal questions in climate can be cast into the paradigm of perturbations and responses as proposed in \cite{Baldovin}. Examples of such questions may in fact be: how much do changes in fresh water fluxes in Antarctica affect sea level rise in the North Atlantic? How do changes in sea surface temperature anomalies in the Pacific Ocean affect temperatures in the Indian Ocean? Answering such questions often relies on quantifying the time-dependent ``flow of information'' along the underlying causal graph rather than discovering the graph itself \cite{Valerio2021,Baldovin} (see also \cite{Ay} in the context of information theory). Such difference with causal discovery methods is further explored and discussed in section \ref{sec:example_null_model}. On the computational side, causal discovery algorithms such as the one based on conditional independence, do not scale to high-dimensional systems \cite{Matt,Camps}. Whereas linear response theory scales to high-dimensional data and allows us to write rigorous, analytical relations between perturbations and responses.\\

It should be noted that linear response theory is an active field of research in climate science \cite{Leith,GhilLucarini,Lucarini2018,Lucarini2017,Lembo,GRITSUN,Majda2010,Pedram1,Pedram2,GritsunIzvestiya,GritsunIzvestiya2}. Such studies, quantifying long-term, forced changes in climate observables, can be broadly grouped in two approaches \cite{Christensen}: the one pioneered by Leith (1975) \cite{Leith}, making use of the fluctuation-dissipation formalism, and the more general formalism proposed by Ruelle (1998) \cite{RUELLE1998220,Ruelle}.\\ 

Our work relates to the approach proposed by Leith (1975) \cite{Leith} and it is based on the recent contribution of Baldovin et al. (2020) \cite{Baldovin}, where the authors presented a clear strategy to infer causality in multivariate linear Markov processes through the fluctuation-dissipation relation. The extension of the proposal of Baldovin et al. (2020) \cite{Baldovin} for studying spatially extended dynamical systems is contingent on two important steps: (i) a methodology to reduce the dimensionality of the system and (ii) a framework for uncertainty estimation. Point (ii) is particularly important when inferring results from real-world data.\\

In this paper, we contribute to (a) dimensionality reduction, (b) linear response theory and (c) causality in climate in the following ways:

\begin{enumerate}[label=\alph*)]
    \item We introduce a scalable computational strategy to decompose a large spatiotemporal climate field into a set of a few \textit{regionally constrained} modes. The average time series inside each pattern quantifies the climate variability of specific regions around the world. The time-dependent linkages among such patterns are then inferred through the fluctuation-dissipation relation. This step allows us to explore how \textit{local} (i.e. regional) variability can influence \textit{remote} locations.
    \item We propose an analytical \textit{null} model for the fluctuation-dissipation relation. The model assigns confidence bounds to the estimated linear responses, therefore distinguishing between \textit{true} and \textit{spurious} results. The proposed strategy allows us for trustworthy statistical inference from real-world data. The application of this model is general and not limited to climate applications.
    \item We showcase the proposed framework on the monthly sea surface temperature (SST) field at global scale. For this step, we consider a 300 years long, stationary integration of a global coupled climate model. Long-distance linkages in the SST field have been characterized in many previous studies. It therefore offers a good real-world test-bed for the methodology. We show how the proposed framework simplifies the description of such a complex, high-dimensional system in an interpretable and comprehensive way.
\end{enumerate}

The paper is organized as follows: in Sec. \ref{sec:framework} we introduce the proposed framework. The data analyzed are described in Sec. \ref{sec:data}. The methodology is applied to climate data in Sec. \ref{sec:climate}. Sec. \ref{sec:conclusions} concludes the work.

\section{Framework} \label{sec:framework}


\subsection{Partitioning climate fields into regionally constrained patterns} \label{sec:dim_red}

Spatiotemporal chaotic fields can be viewed as dynamical systems $\boldsymbol{x}(t) \in \mathbb{R}^{N}$ living in a $N$-dimensional state space \cite{Gibson,Predrag}. 
The dimensionality $N$ is theoretically infinite but in practice equal to the number of grid cells used to discretize the longitude, latitude and vertical coordinates (times the total number of variables; e.g., temperature, velocities etc.) \cite{PRX}. In the case of dissipative chaotic systems, such high-dimensional dynamics is confined on lower-dimensional objects known as ``inertial manifolds'' or ``attractors'' \cite{Predrag,DingPredrag,FarandaSciRep}. The \textit{effective} dimensionality of the system \cite{theiler} is then finite and given by the attractor dimension $D$. This is arguably the case of large scale climate dynamics, where recurrent spatiotemporal patterns, known as modes of variability (e.g., ENSO, monsoon system, Indian Ocean modes \cite{Klein,Webster,Falasca} etc.) are a manifestation of the low dimensionality of the climate attractor \cite{Dubrulle,PRX}.\\

Here the goal is to coarse grain a $N$-dimensional climate field into a set of very few (order 10) patterns. Crucially, such components should be \textit{regionally constrained} in longitude-latitude space. Requiring for the identification of regional patterns is a desirable property as climate variability can be often thought of as a set of \textit{remote} responses driven by \textit{local} perturbations.
A clear example is given by the climate system's response to El Ni\~no events \cite{ENSOcomplexity}. An El Ni\~no phenomenon is characterized by a build-up of warm sea surface temperature in the eastern Pacific. Such local warming excites an atmospheric wave response resulting in the heating of the whole tropical troposphere \cite{Gill,Sobel}. As a result, an El Ni\~no episode in the Pacific causes a warming in both the Indian and tropical Atlantic basins at later times. Therefore, when reducing the dimensionality of the climate system it is useful to distinguish between climate phenomena driven by local dynamics or forced by remote variability. In this paper, we will do so by first coarse graining the system in terms of regional modes of variability. At a second step, we will infer their \textit{causal} linkages via linear response theory \cite{Baldovin}.\\

In this section we show that adding a simple constraint to community detection methodologies \cite{Lancichinetti,Barabasi} provides a scalable and practical framework to identify regionally constrained modes of variability in climate fields. The strategy proposed here is based on two main steps: first, given a field saved as a data matrix $\boldsymbol{x} \in \mathbb{R}^{N,T}$ we infer a graph between its $N$ time series based on both their covariability and distance. We then identify communities in such graph, thus partitioning the original data into a few components. Communities will consist of sets of highly correlated time series and will serve as proxies of climate modes of variability. In Appendix \ref{app:dim_red_methods} we discuss strengths and limitations of current dimensionality reduction methods and further motivate our proposal.\\

In practice, in this work $\boldsymbol{x} \in \mathbb{R}^{N,T}$ will be specified by the sea surface temperature field only. Components of the field $x_{i}(t)$ will then represent a sea surface temperature time series at grid point $i$. $N$ will be the number of grid points and $T$ the length of each time series at a given temporal resolution. The framework proposed is however general and can work with multivariate fields.

\subsubsection{Graph inference} \label{sec:graph_inference}
Consider a spatiotemporal field saved as a data matrix $\boldsymbol{x} \in \mathbb{R}^{N,T}$, with $N$ time series of length $T$. Given a pair of time series $x_{i}(t)$ and $x_{j}(t)$, scaled to zero mean, we compute their covariance at lag $\tau = 0$, $C_{i,j} = \overline{x_{i}(t) x_{j}(t)}$; where $\overline{f}$ stands for the temporal average of function $f$. An undirected, unweighted graph can then be encoded in an adjacency matrix $\boldsymbol{A} \in \mathbb{R}^{\textit{N,N}}$ as:
\begin{linenomath}
\begin{equation}
A_{i,j} = 
\begin{cases}
1 - \delta_{i,j} &  ~ \text{if} ~ C_{i,j} \geq k~ \text{and} ~ d(i,j) \leq \eta\\
0 &  ~ \text{otherwise}
\end{cases}
\label{eq:adjacency_matrix}
\end{equation}
\end{linenomath}
Where the Kronecker delta $\delta_{i,j}$ allows us to remove ``self-links''. The parameter $k$ sets the minimum covariance that two time series must have to be connected. The parameter $d(i,j)$ is the distance between grid cells $i$ and $j$, and $\eta$ is a distance threshold. The rationale behind this choice is that we consider two time series $x_i(t)$ and $x_j(t)$ linked to each other if (a) their covariance is larger than a threshold $k$ and (b) if they are relatively close in the spatial domain considered. Importantly, $d(i,j)$ is computed using the Haversine (or great-circle) distance, determining the angular distance between two points $i$ and $j$ on a sphere as a function of their longitudes and latitudes \cite{scikit-learn}.\\ 

Both thresholds $k$ and $\eta$ can be specified by the user. However, their optimal values will largely depend on the statistics of the field of interest (e.g., sea surface temperature, cloud fraction) and by the spatial domain considered (e.g., regional or global domains). We therefore propose two heuristics to compute such parameters as a function of the data matrix $\boldsymbol{x} \in \mathbb{R}^{N,T}$.\\

\paragraph{Heuristic for parameter $k$.} Given time series $x_{i}(t)$ and $x_{j}(t)$: (a) compute covariances $C_{i,j}$, $\forall i,j; i \neq j$ and (b) set $k$ as a high quantile $q_k$ of the distribution of all covariances $C_{i,j}$. To make this idea feasible in practice, we can approximate such distribution by random sampling $S_k$ pairs of time series $x_{i}(t)$ and $x_{j}(t)$ and then computing their covariances. $k$ is then estimated as a high quantile $q_k$ of the sampled distribution. A pragmatic choice of $q_k$ is $q_k = 0.95$ as we observed in different applications that is a good compromise between the identification of a sparse, but not too sparse, graph. The sampling size considered here is $S_k = 10^6$.\\

\paragraph{Heuristic for parameter $\eta$.} Given time series $x_{i}(t)$ and $x_{j}(t)$ embedded at grid points $i$ and $j$ (a) calculate the Haversine distance $d(i,j)$ and (b) estimate $\eta$ as a low quantile $q_{\eta}$ of the distribution of all distances $d(i,j)$. As for the parameter $k$, in practice the distribution of distances can be approximated by random sampling $S_{\eta}$ pairs of locations $i$ and $j$ and computing their Haversine distance. We choose $q_{\eta} = 0.15$, with no large sensitivity over such value, and $S_{\eta} = 10^6$.\\

\subsubsection{Detecting patterns} \label{sec:community_detection}
Sets of highly correlated time series (i.e., modes) in the original field $\boldsymbol{x} \in \mathbb{R}^{N,T}$  correspond to groups of nodes that are more interconnected to each other than to the rest of the graph, in other words ``communities'' \cite{Barabasi}. Fast and scalable community detection algorithms \cite{Lancichinetti} can be leveraged to reduce the dimensionality of the graph in Eq. \ref{eq:adjacency_matrix}. In this study, we consider the Infomap methodology \cite{Rosvall1,Rosvall2}. Such method is based on the Map Equation \cite{MapEq,RosvallReview} and casts the problem of community detection as an optimal compression problem \cite{Rosvall2}. Mainly, Infomap exploits the community structure to minimize the description of a random walk on the graph \cite{MapEq}. Such methodology has been found to be the best performing community detection in different benchmarks \cite{Lancichinetti}, and has shown excellent performance in a previous climate study \cite{Tantet}. In what follows we are going to refer to the identified communities as ``patterns'', ``modes'' or ``communities'' interchangeably.\\

The number and size of the identified patterns will depend on parameters $q_k$ and $q_{\eta}$ introduced in section \ref{sec:graph_inference}. A priori knowledge of the system can help setting the values of parameters $q_k$ and $q_{\eta}$. As a rule of thumb, we recommend the interested practitioner to first gain an intuition about the underlying network structure of the system by setting  $q_k = 0.95$ but without constraints on distances (this corresponds to setting $q_{\eta} = 1$). The $q_{\eta}$ parameter is then used to remove long-distance (in longitude-latitude space) dependencies. We recommend starting from $q_{\eta} = 0.15$ as in this study. If the resulting patterns are still not regionally constrained, we suggest to lower $q_{\eta}$ to slightly smaller values, e.g. $q_{\eta} = 0.1$. Finally, we note that in different applications, ranging from reducing the dimensionality of sea level to outgoing longwave radiation fields, we found no need to change the values of parameters $q_k = 0.95$ and $q_{\eta} = 0.15$.


\subsubsection{Defining signals (time series)} \label{sec:signals}
Given a set of $n$ patterns $c = (c_1,c_2,c_3,...c_n)$ we study their temporal variability as the average time series inside. Formally, for each $c_{j}$ we can define its respective time series as: 
\begin{linenomath}
\begin{equation}
X(c_{j},t) = \frac{1}{\sum_{i \in c_{j}}\cos(\theta_{i})} \sum_{i \in c_{j}} x_{i}(t) \cos(\theta_{i}),
\label{eq:signals}
\end{equation}
\end{linenomath}
where $\theta_{i}$ is the latitude of $x_{i}(t)$. The term $\cos(\theta_{i})$ allows us to
implement the area-weighted averaging on a uniform longitude-latitude grid. We note that another way to define each signal is the area-integrated anomaly $X(c_{j},t) = \sum_{i \in c_{j}} x_{i}(t) \cos(\theta_{i})$. Such definition allows us to rank different patterns $c_j$ with respect to both their variability and their size, therefore carrying different weights in the linear response formulas. The definition of signals through area-integrated anomalies can be useful in climate change experiments performed with linear response theory, and it will be considered in future studies. In this work we adopt the definition in Eq. \ref{eq:signals}.\\

In this study, the graph inference step in Eq. \ref{eq:adjacency_matrix} considers correlations rather than covariances, therefore $C_{i,j} = \overline{x_{i}(t) x_{j}(t)}$ in Eq. \ref{eq:adjacency_matrix} are computed after scaling $x_{i}(t)$ and $x_{j}(t)$ to zero mean and unit variance. We used correlations for qualitative comparison with the $\delta$-MAPS framework \cite{Ilias_1,dimRedClimate}, but covariances can be considered in future work.\\

\subsection{Linear response theory and fluctuation-dissipation relation} \label{sec:FDR}

Baldovin et al. (2020) \cite{Baldovin}, proposed the following physical definition of causality: given a dynamical system $\boldsymbol{x}(t) = [x_{1}(t),x_{2}(t),...,x_{N}(t)]$ with $N$ time series, each of length $T$ we say that $x_{j}$ causes $x_{k}$, i.e. $x_{j} \rightarrow x_{k}$,  if a small perturbation applied to variable $x_{j}$ at time $t = 0$, i.e. $x_{j}(0) + \delta x_{j}(0)$, induces \textit{on average} a change on variable $x_{k}(\tau)$ at a later time $t = \tau$. We note that the contribution of Lucarini, V. (2018) \cite{Lucarini2018} pursues close scientific goals to the work of Baldovin et al. (2021) \cite{Baldovin}.

\subsubsection{General case}

Consider a Markov process $\boldsymbol{x}(t) = [x_{1}(t),x_{2}(t),...,x_{N}(t)]$. Each time series $x_{i}(t)$ is scaled to zero mean. The system is stationary with invariant probability distribution $\rho(\boldsymbol{x})$. We perturb the system $\boldsymbol{x}(t)$ at time $t = 0$ with a small, impulse perturbation $\delta\boldsymbol{x}(0) = [\delta x_{1}(0),\delta x_{2}(0),...,\delta x_{N}(0)]$. We aim to answer the following question: how does this \textit{external} perturbation $\delta\boldsymbol{x}(0)$ affect the whole system $\boldsymbol{x}(\tau)$ at time $t = \tau$, on average? Formally, we are interested in quantifying the following object:
\begin{linenomath}
\begin{equation}
\delta \langle x_{k}(\tau) \rangle = \langle x_{k}(\tau) \rangle_{p} - \langle x_{k}(\tau) \rangle,
\label{eq:different}
\end{equation}
\end{linenomath}
where the brackets $\langle x_{k}(\tau) \rangle$ indicate the ensemble averages of $x_{k}(\tau)$, i.e. the average over many realizations of the system, and the subscript $p$ specifies the perturbed dynamics. Therefore, Eq. \ref{eq:different} defines the difference between the components $x_{k}(\tau)$ of the perturbed and unperturbed systems in the \textit{average} sense. Eq. \ref{eq:different} can be used to study changes $\delta \langle O(x_{k}(\tau)) \rangle$ of a generic observable $O(x_{k}(\tau))$ (i.e., a measurable quantity, function of the state space vector $\boldsymbol{x}(\tau)$ at time $t = \tau$). To study causality, here we simply consider the identity case $O(x_{k}(\tau)) = x_{k}(\tau)$, see  \cite{Baldovin}.\\

Under the assumption of a small perturbation $\delta\boldsymbol{x}(0)$ and with $\rho(\boldsymbol{x})$ sufficiently smooth and non-vanishing, the following result holds:
\begin{linenomath}
\begin{equation}
R_{k,j}(\tau) = \lim_{\delta x_{j}(0)\to0} \frac{\delta \langle x_{k}(\tau) \rangle}{\delta x_{j}(0)} = - \Big\langle x_{k}(\tau) \frac{\partial \ln \rho(\boldsymbol{x})}{\partial x_{j}} \Big|_{\boldsymbol{x}(0)} \Big\rangle .
\label{eq:response_general} 
\end{equation}
\end{linenomath}
$\boldsymbol{R}(\tau)$ is the linear response matrix and we refer to section II of Boffetta et al. (2003) \cite{boffett} for a derivation of Eq. \ref{eq:response_general}. $R_{k,j}(\tau)$ quantifies the response of a variable $x_{k}(\tau)$ at time $t = \tau$ given a small perturbation $\delta x_{j}(0)$ applied to variable $x_{j}(0)$ at time $t = 0$.  Eq. \ref{eq:response_general} is known as the generalized fluctuation-dissipation relation (FDR) and valid for both linear and nonlinear systems \cite{Marconi}. Note that in case of deterministic systems the invariant measure $\rho(\boldsymbol{x})$ is singular almost everywhere on the attractor. Therefore in practice one needs to add Gaussian noise even to deterministic systems in order to ``smooth'' the probability distribution before applying FDR as proposed here \cite{GRITSUN}.\\

Eq. \ref{eq:response_general} is a powerful formula as it allows us to compute responses to perturbations solely given the gradients of the probability distribution $\rho(\boldsymbol{x})$ of the \textit{unperturbed} system. However, the functional form of $\rho(\boldsymbol{x})$ is not known a priori and can be highly nontrivial, especially for high-dimensional systems. To overcome such issue, applications often focus on the simpler case of Gaussian distributions (see for example \cite{Leith,Christensen}). This is the case of linear systems as shown in the next section.

\subsubsection{Linear systems and quasi-Gaussian approximation} \label{sec:quasiGaussian}

We now consider a $N$ dimensional stochastic linear process $\boldsymbol{x}(t) = [x_{1}(t),x_{2}(t),...,x_{N}(t)]$ governed by the following equation:
\begin{linenomath}
\begin{equation}
\boldsymbol{x}(t+1) = \boldsymbol{M} \boldsymbol{x}(t) +\boldsymbol{B} \boldsymbol{\xi}(t).
\label{eq:linear_markov}
\end{equation}
\end{linenomath}
The matrix $\boldsymbol{M} \in \mathbb{R}^{N,N}$ specifies the deterministic dynamics of the system. The term $\boldsymbol{\xi}(t) \in \mathbb{R}^{N}$ with $\xi_{i}(t) \overset{\mathrm{iid}}{\sim} \mathcal{N}(0,1)$ represents a delta correlated white noise (i.e., $\langle \xi(t) \xi(s) \rangle = \delta_{t,s}$). The matrix $\boldsymbol{B} \in \mathbb{R}^{N,N}$ specifies the amplitude of the noise (i.e., standard deviation). The probability distribution $\rho(\boldsymbol{x})$ is Gaussian and Eq. \ref{eq:response_general} factorizes to:
\begin{linenomath}
\begin{equation}
\boldsymbol{R}(\tau) = \boldsymbol{M}^{\tau} = \boldsymbol{C}(\tau)\boldsymbol{C}(0)^{-1}.
\label{eq:response_linear_markov} 
\end{equation}
\end{linenomath}
Where the covariance function $C_{i,j}(\tau) = \langle x_{i}(t+\tau) x_{j}(t) \rangle$ ($x_{i}$ is assumed to be zero mean).
Eq. \ref{eq:response_linear_markov} shows that the response of a linear system to small \textit{external} perturbations is a function of covariance matrices computed for the stationary (i.e., unperturbed) dynamics \cite{boffett}.\\

\paragraph*{Relevance for nonlinear systems.} The form of FDR shown in Eq. \ref{eq:response_linear_markov} has been the one commonly used in climate applications and it is commonly referred to as ``quasi-Gaussian approximation'' \cite{GRITSUN,Ring,Majda2010,Pedram1,Pedram2}. Importantly, it has been shown that such formula performs well for weakly nonlinear systems. For instance Baldovin et al. (2020) \cite{Baldovin} showed remarkably good results when analyzing linear responses in a Langevin equation with a quartic potential. Gritsun et al. (2007) \cite{GRITSUN} also pointed out how this formula works well for non-Gaussian systems with second order nonlinearities. Additionally, Eq. \ref{eq:response_linear_markov} has been shown to give reliable results in the case of nonlinear deterministic dynamical systems also in case of finite perturbations, see Fig. 1 in Boffetta et al. (2003) \cite{boffett}. Furthermore, we will show in Appendix \ref{app:histograms_global} that the probability distributions considered in this study can be well approximated by Gaussians, further justifying the use of this approximation in our context.\\

Results presented in this section hold in the sense of ensemble average, therefore covariance matrices $\boldsymbol{C}(\tau)$ and $\boldsymbol{C}(0)$ are computed by averaging over many realizations of the system. The computation of ensemble averages gives rise to an additional complication in real-world data for which we only have access to a single trajectory.

\subsection{A \textit{null model} for the fluctuation-dissipation relation} \label{sec:FDR_null_model}
In real-world applications we cannot compute ensemble averages. The common way to overcome this problem and reconcile data analysis with theory, is through the assumption of ergodicity \cite{nonEqStatMech}. If the system $\boldsymbol{x}$ is ergodic it holds: $\overline{O(\boldsymbol{x})} = \langle O(\boldsymbol{x}) \rangle$ in the limit  $T \rightarrow \infty$; where $O(\boldsymbol{x})$ is a general observable, $\overline{O(\boldsymbol{x})}$ indicates the time average and $T$ is the length of the trajectory $\boldsymbol{x}$.\\

Ergodicity is the main assumption behind any climate study using the fluctuation-dissipation theorem (see \cite{Pedram1} and references therein). Covariance matrices are then estimated using temporal averages, e.g. $C_{i,j}(\tau) = \overline{x_{i}(t+\tau) x_{j}(t)}$ ($x_{i}$ is assumed to be zero mean), but again we are left with the problem of observing the system over a finite time window. Therefore we can always expect \textit{spurious} results when estimating response functions. Spurious results come from two main contributors: (a) finite samples (i.e., the length $T$ of the trajectory is finite) and (b) large autocorrelations of the underlying time series $x_i(t)$.\\

To the best of our knowledge, an analytical statistical test to distinguish between \textit{spurious} and \textit{real} responses in the linear response theory formalism has not been proposed in the literature. Here we fill this gap by proposing a \textit{null} model for fluctuation-dissipation relation and derive its analytical solution. We start by proposing a null hypothesis for a general stochastic dynamical system.\\

\paragraph{Null hypothesis.} Given a system $\boldsymbol{x}(t) = [x_{1}(t),x_{2}(t),...,x_{N}(t)]$ it holds $R_{k,j}(\tau) = 0, ~ \forall j,k = 1,...,N; ~ \text{with} ~ j \neq k$. In the context of causality this implies that there is no causal link $x_{j} \rightarrow x_{k}, ~ \forall j,k = 1,...,N; j \neq k$. \\

\paragraph{Null model.} Given a process saved as a data matrix $\boldsymbol{x} \in \mathbb{R}^{N,T}$, we define a new process $\boldsymbol{\Tilde{x}} \in \mathbb{R}^{N,T}$ simulated by a null model. Every time series in $\boldsymbol{x}$ and $\boldsymbol{\Tilde{x}}$ are rescaled to zero mean. The null model takes the following form: 
\begin{linenomath}
\begin{equation}
\begin{split}
\boldsymbol{\Tilde{x}}(t+1) &= \boldsymbol{\Tilde{M}} \boldsymbol{\Tilde{x}}(t) +\boldsymbol{\Tilde{B}} \boldsymbol{\xi}(t) \\
\small{\textit{with}} ~ \boldsymbol{\Tilde{M}} &= 
 \begin{pmatrix}
  \phi_{1}             & 0               & \cdots & 0 \\
  0                    & \phi_{2}      & \cdots & 0 \\
  \vdots               & \vdots               & \ddots & \vdots  \\
  0                    & 0                    & \cdots & \phi_{N}
 \end{pmatrix} ; \\
 \boldsymbol{\Tilde{B}} &= 
 \begin{pmatrix}
\Tilde{\sigma}_{1}       & 0       & \cdots & 0 \\
0                  & \Tilde{\sigma}_{2} & \cdots & 0 \\
\vdots             & \vdots       & \ddots & \vdots  \\
0                  & 0            & \cdots & \Tilde{\sigma}_{N}
\end{pmatrix};\\
  \xi_{i}(t) &\overset{\mathrm{iid}}{\sim} \mathcal{N}(0,1), ~ i = 1,...,N.
\end{split}
\label{eq:linear_markov_matrix_red_noise}
\end{equation}
\end{linenomath}
Here, $\phi_{i}$ is the lag-1 autocorrelation of the ``original'' time series $x_{i}(t)$; $\Tilde{\sigma}_{i} = \sigma_{i} (1 - \phi_{i}^{2})$ is the standard deviation of the Gaussian noise, where $\sigma_i$ is the standard deviation of the ``original'' time series $x_{i}(t)$. Therefore, each time series $\Tilde{x}_{i}(t)$ has the same mean, variance and lag-1 autocorrelation of $x_{i}(t)$, however every pair $\Tilde{x}_{i}(t)$, $\Tilde{x}_{j}(t)$ is now independent. Note that the null model in Eq. \ref{eq:linear_markov_matrix_red_noise} is largely inspired by the commonly adopted red noise test in climate analysis \cite{GhilChildress,Imkeller,AllenSmith,Henk_1}.\\

The matrix $\boldsymbol{\Tilde{M}}$, defining the deterministic evolution, is diagonal; therefore at asymptotic times $T \rightarrow \infty$ there is no causal link among variables. However, for finite time windows, the response matrix estimated through time averaged covariance matrices as $\boldsymbol{R}(\tau) = \boldsymbol{C}(\tau)\boldsymbol{C}(0)^{-1}$ will give rise to \textit{spurious} off-diagonal elements. The distribution of responses of the null process $\boldsymbol{\Tilde{x}}$ defines confidence bounds of responses of the original process $\boldsymbol{x}$.\\

To compute the confidence level of the response $R_{k,j}(\tau)$ at each lag $\tau$ we first propose a numerical implementation. We then solve the problem analytically for the case $T\gg1$.

\subsubsection{Confidence bounds of the response matrix: numerical estimation} \label{sec:numerics_null_model}

Given a field $\boldsymbol{x} \in \mathbb{R}^{N,T}$, our goal is to provide an estimation of confidence intervals of the response matrix $\boldsymbol{R}(\tau)$ at each lag $\tau$, with $\tau = 0,1,...,\tau_{\infty}$. Such bounds can be numerically estimated as follows: 

\begin{enumerate}[label=\roman*)]
    \item Generate a new process $\boldsymbol{\Tilde{x}} \in \mathbb{R}^{N,T}$ using the null model proposed in Eq. \ref{eq:linear_markov_matrix_red_noise}.
    \item Estimate the response matrix $\boldsymbol{R}(\tau)$ of the null model $\boldsymbol{\Tilde{x}}(t)$ for lags $\tau \in [0,\tau_{\infty}]$.
    \item Repeat the two steps above for $B$ times, ($B$ should be large, $B \gg 1$), therefore creating an ensemble of \textit{null} responses.
    \item For each lag $\tau$ we obtain a distribution of possible responses generated by the null model. Confidence bounds of responses can be estimated as low and high quantiles of the distribution, or as chosen in this paper, multiples of its standard deviation.
\end{enumerate}

\subsubsection{Confidence bounds of the response matrix: analytical derivation} \label{sec:analytical_null_model}

We note that the analytical form of the response matrix in the null model in Eq. \ref{eq:linear_markov_matrix_red_noise} is trivial and given by $\boldsymbol{R}(\tau) = \boldsymbol{M}^{\tau}$ with entries $\phi_{k}^{\tau} \delta_{k,j}$; $\delta_{k,j}$ being the Kronecker delta. However, estimating responses from time series of finite length, will give rise to spurious results departing from the expected value of $\boldsymbol{M}^{\tau}$.\\

In this section we present the analytical form of the probability distribution of responses \textit{estimated} by the formula $\boldsymbol{R}(\tau) = \boldsymbol{C}(\tau)\boldsymbol{C}(0)^{-1}$ in the case of time series generated by the \textit{null} model in Eq. \ref{eq:linear_markov_matrix_red_noise}. We then refer the reader to Appendix \ref{app:confidence_bounds} for the derivation.\\ 

The main assumption is that \textit{null} responses $R_{k,j}(\tau)$ follow a Normal distribution. Therefore the expected value $\mathbb{E}[R_{k,j}(\tau)] = \langle R_{k,j}(\tau) \rangle$ and variance $\mathbb{V}\textrm{ar}[R_{k,j}(\tau)] = \langle (R_{k,j}(\tau) - \langle R_{k,j}(\tau) \rangle )^{2} \rangle$ uniquely define the probability distribution $\rho(R_{k,j}(\tau))$. We have:
\begin{linenomath}
\begin{equation}
\begin{split}
\mathbb{E}[R_{k,j}(\tau)] &= \phi_{k}^{\tau} \delta_{k,j}\\
\mathbb{V}\textrm{ar}[R_{k,j}(\tau)] &= \frac{\phi_k^{2\tau}-1}{T}+\frac{2}{T} \Big{(}\frac{1-\phi_k^{\tau}\phi_j^{\tau}}{1-\phi_k\phi_j} \Big{)}\\ 
&- \frac{2\phi_{k}^{\tau}}{T} \Big{(} \phi_{k} \frac{\phi_{j}^{\tau}-\phi_{k}^{\tau}}{\phi_{j}-\phi_{k}} \Big{)} ~ .
\end{split}
\label{eq:expectation_and_Variance_R}
\end{equation}
\end{linenomath}
Finally, in the case $\phi_{k} = \phi_{j}$ we substitute the term $\phi_{k}\frac{\phi_{j}^{\tau}-\phi_{k}^{\tau}}{\phi_{j}-\phi_{k}}$ with the limit:
\begin{linenomath}
\begin{equation}
\lim_{\phi_{j}\to\phi_{k}} \phi_{k} \frac{\phi_{j}^{\tau} - \phi_{k}^{\tau}}{\phi_{j} - \phi_{k}} = \phi_{k}^{\tau} \tau.
\label{eq:limit} 
\end{equation}
\end{linenomath}

Equation \ref{eq:expectation_and_Variance_R} assumes that each time series has been previously normalized to zero mean and unit variance. In case of non-standardized time series, Eq. \ref{eq:expectation_and_Variance_R} becomes $(\sigma^{2}_k/\sigma^{2}_j) \times \text{Eq. \ref{eq:expectation_and_Variance_R}}$; $\sigma^{2}_i$ being the variance of time series $x_{i}(t)$ (see also Eq. 15 in \cite{Baldovin}).\\

In this paper, confidence bounds are always defined by $\mathbb{E}[R_{k,j}(\tau)] \pm 3 \sqrt{\mathbb{V}\textrm{ar}[R_{k,j}(\tau)]}$ (i.e., $\pm 3 \sigma$ confidence level).\\

Finally, we note that the analytical confidence bounds proposed in Eq. \ref{eq:expectation_and_Variance_R} overcome an important problem in climate applications of linear response theory. Previous studies such as \cite{GRITSUN,Majda2010,Pedram2} focused on evaluating the integral $\int_{0}^{\infty} d\tau ~ \boldsymbol{R}(\tau)$. In practice, the upper bound of the integral needs to be specified by a $\tau_{\infty}$ much larger than the characteristic time of the response. However, responses at larger lags are affected by spurious results and $\tau_{\infty}$ has been set to values as low as $30$ days in some studies (e.g., \cite{GRITSUN,GritsunMajda}) or it has been tuned to have the best performance of FDR in  others \cite{Pedram2}. The confidence bounds proposed in this section can then be leveraged to neglect spurious terms, study responses at longer time scales and obtain results largely independent of $\tau_{\infty}$.

\subsection{A simple example} \label{sec:example_null_model}

\begin{figure}[tbhp]
\centering
\includegraphics[width=1\linewidth]{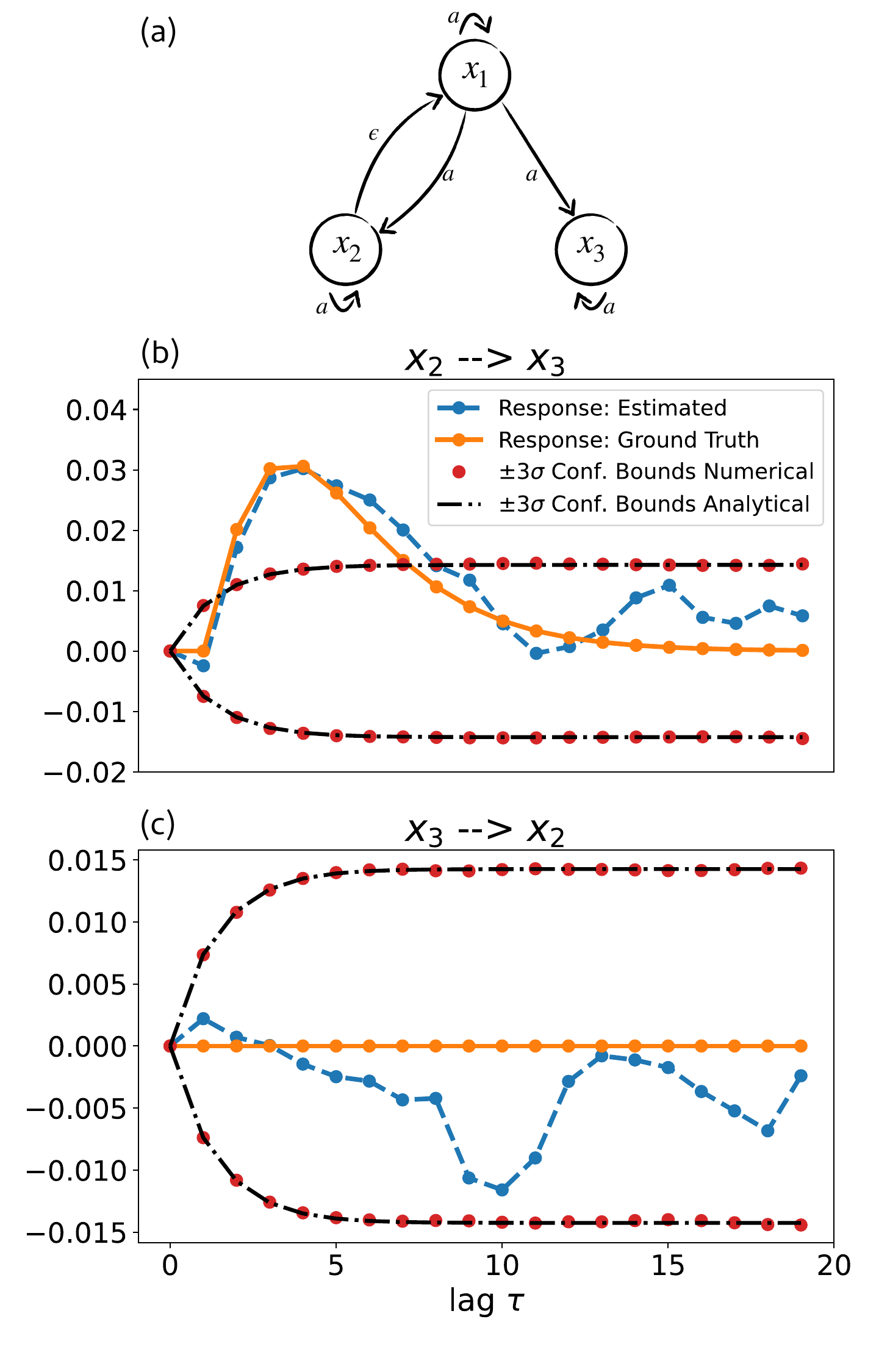}
\caption{Panel (a): Graph representing the Markov model in Eq. \ref{eq:linear_markov_matrix_example}. This is the same simple system considered in Baldovin et al. (2020) \cite{Baldovin} (where here $[x_{1},x_{2},x_{3}]$ correspond to their $[x,y,z]$ in \cite{Baldovin}). Panel (b): response of variable $x_{3}$ when perturbing $x_{2}$, i.e. testing for link $x_{2} \rightarrow x_{3}$. Panel (c): response of variable $x_{2}$ when perturbing $x_{3}$, i.e. testing for link $x_{3} \rightarrow x_{2}$. All time series have been rescaled to zero mean and unit variance before computing responses. The response ground truths are shown as solid orange lines. Dashed blue lines are responses estimated through temporal averages: for this step we use a long trajectory of length $T = 10^5$ simulated by system in Eq. \ref{eq:linear_markov_matrix_example}. Red dots indicate the confidence bounds computed numerically using $B = 10^4$ ensemble members of the \textit{null} model as shown in \ref{sec:numerics_null_model}. In each panel, the dot-dashed black line is the analytical solution as in Eq. \ref{eq:expectation_and_Variance_R}. Bounds correspond to the $\pm 3 \sigma$ confidence level. All estimated responses (i.e. blue curves) in between the confidence bounds are here considered as spurious.}
\label{fig:test_temporal_average}
\end{figure}

We test the confidence bounds proposed in section \ref{sec:FDR_null_model} in the context of a simple linear Markov model. We choose the same test model used in \cite{Baldovin} in order to compare results and show differences between approaches. The system considered is the following:
\begin{linenomath}
\begin{equation}
\begin{split}
\boldsymbol{x}(t+1) &= \boldsymbol{M} \boldsymbol{x}(t) +\boldsymbol{B} \boldsymbol{\xi}(t) \\
\small{\textit{with}} ~ \boldsymbol{M} &= 
 \begin{pmatrix}
  a & \epsilon & 0 \\
  a & a & 0 \\
  a & 0 & a
 \end{pmatrix} ; \\
 \boldsymbol{B} &= 
 \begin{pmatrix}
  b & 0 & 0\\
  0 & b & 0 \\
  0 & 0 & b
 \end{pmatrix};\\
 \xi_{i}(t) &\overset{\mathrm{iid}}{\sim} \mathcal{N}(0,1), ~ i = 1,2,3.
\end{split}
\label{eq:linear_markov_matrix_example}
\end{equation}
\end{linenomath}
As in \cite{Baldovin}, we set $a = 0.5$ and $b = 1$; we then set $\epsilon = 0.04$. Note that here $[x_{1},x_{2},x_{3}]$ correspond to $[x,y,z]$ in \cite{Baldovin}. In this simple model, a small perturbation applied on variable $x_{2}$ would propagate through the system and cause a change first at variable $x_{1}$ and then at $x_{3}$ \cite{Baldovin}. However, a perturbation in $x_{3}$ cannot reach either $x_{1}$ or $x_{2}$; this is clear by looking at the underlying graph in Fig. \ref{fig:test_temporal_average}(a). We first focus on the true responses $\boldsymbol{R}_{k,j}(\tau)$, here computed as $\boldsymbol{R}_{k,j}(\tau) = \frac{\sigma_j}{\sigma_k} [\boldsymbol{M}^{\tau}]_{k,j}$  and shown in Fig. \ref{fig:test_temporal_average}. The links $x_{2} \rightarrow x_{3}$ (i.e, $R_{3,2}(\tau)$) and $x_{3} \rightarrow x_{2}$ (i.e, $R_{2,3}(\tau)$) are correctly captured: the first nonzero response $R_{3,2}(\tau)$ is identified at lag $\tau = 2$ and responses $R_{2,3}(\tau)$ are found to be zero for any lag $\tau$ (see Fig. \ref{fig:test_temporal_average}(b,c)). 
Note that such results cannot be inferred with correlations only. For example, the estimation of the link $x_{3} \rightarrow x_{2}$ via correlations will give non-zero values because of the confounder $x_{1}$. We refer to Baldovin et al. (2020) \cite{Baldovin} for a thorough comparison with correlation analysis.\\ 

Let us briefly note here the main conceptual difference between the fluctuation-response formalism and methods for causal discovery. Causal discovery methods used in climate and based on conditional independence such as \cite{RungeScience} aim in discovering the underlying causal graph in Fig. \ref{fig:test_temporal_average}(a) given time series data. Therefore, the link $x_{2} \rightarrow x_{3}$ would not be identified as a causal link. The same holds for Granger causality and transfer entropy \cite{Granger,Schreiber} as shown in \cite{Baldovin}. However, in a physical experiment an intervention over variable $x_{2}$ would cause a change in variable $x_{3}$. Such ``interventional'' view of causation is the one considered here and can be correctly captured by linear response theory in a straightforward way, see  Fig. \ref{fig:test_temporal_average}(b). We refer to section IIIA of \cite{Baldovin} for an in-depth discussion.\\ 

In real-world cases we deal with time series of finite length. We then simulate the system in Eq. \ref{eq:linear_markov_matrix_example} for $T = 10^{5}$ time steps and estimate the causal links $x_{j} \rightarrow x_{k}$ using the response formalism $R_{k,j}$ (i.e., formula \ref{eq:response_linear_markov} after standardizing each $x_{i}$ to unit variance)using temporal averages. As expected, in this case our results are affected by spurious terms, see blue dashed lines in Fig. \ref{fig:test_temporal_average}. The null model proposed in Eq. \ref{eq:linear_markov_matrix_red_noise} is then leveraged to assign confidence bounds to the estimated responses. In Fig. \ref{fig:test_temporal_average}b and \ref{fig:test_temporal_average}c, we report both the numerically estimated and analytically derived (Eq. \ref{eq:expectation_and_Variance_R}) confidence bounds. Responses inside the confidence bounds in Fig. \ref{fig:test_temporal_average} can be considered as spurious. The confidence bounds correctly identify the non-zero responses $R_{3,2}(\tau)$ for $\tau = 1$ and large lags as spurious results, see Fig. \ref{fig:test_temporal_average}(b). Additionally, the test allows us to disregard the spurious link $x_{3} \rightarrow x_{2}$, see Fig. \ref{fig:test_temporal_average}(c). Responses $R_{k,j}(\tau)$, i.e. links $x_{j} \rightarrow x_{k}$, and confidence bounds for every $j$ and $k$ are reported in Appendix \ref{sec:conf_bounds_test}.

\subsection{Metrics} \label{sec:metrics}

The framework allows us to identify any causal interaction $x_{j} \rightarrow x_{k}$ given the definition of causality presented in \cite{Baldovin}. Given $N$ time series this means $N^{2}$ links at each time-lag $\tau$. Analyzing all interactions in such network gets infeasible with larger $N$. We then introduce a few metrics to analyze such causal graphs. In \cite{Baldovin}, the authors proposed a simple ``cumulative degree of causation'' of each link $x_{j} \rightarrow x_{k}$ as a Kubo formula \cite{KuboBook}. Here we consider the same formula while summing over the statistically significant responses $R_{k,j}(\tau^{*})$, defined at lags $\tau^{*}$. We compute respones $R_{k,j}(\tau)$ up to a maximum lag $\tau_{\infty}$; theoretically, the summation would be up to $\infty$, in practice we choose a $\tau_{\infty}$ much longer than the characteristic time of the response. The ``cumulative degree of causation'' considered here is then defined as follows:
\begin{linenomath}
\begin{equation}
\mathcal{D}_{j \rightarrow k} = \sum_{\tau^{*}}^{\tau_{\infty}} R_{k,j}(\tau^{*}) \\
\label{eq:causal_strength_j_to_k_Kubo}
\end{equation}
\end{linenomath}
Since responses can be negative and positive, the degree of causation such as in Eq. \ref{eq:causal_strength_j_to_k_Kubo} can be zero even in the presence of causal links. It can be therefore useful to consider a modified version of Eq. \ref{eq:causal_strength_j_to_k_Kubo} by summing over the absolute value of responses as follows:
\begin{linenomath}
\begin{equation}
\mathcal{D}^{*}_{j \rightarrow k} = \sum_{\tau^{*}}^{\tau_{\infty}} \mid R_{k,j}(\tau^{*}) \mid \\
\label{eq:causal_strength_j_to_k}
\end{equation}
\end{linenomath}

Equations \ref{eq:causal_strength_j_to_k_Kubo} and \ref{eq:causal_strength_j_to_k} quantify the cumulative response of any variables $x_k$ to perturbations at $x_j$, i.e. the ``strength'' of the causal link $x_j \rightarrow x_k$.\\

Finally, we rank each variable $x_j$ by defining its ``causal strength'' as follows:
\begin{linenomath}
\begin{equation}
\mathcal{D}_{j} = \sum_{k = 1}^{N} \mathcal{D}^{*}_{j \rightarrow k}~; ~ j \neq k
\label{eq:total_causal_strength_j}
\end{equation}
\end{linenomath}
Eq. \ref{eq:total_causal_strength_j} allows us to rank nodes in the climate network in regards to their ability to \textit{causally} influence other nodes. Informally, large values of $\mathcal{D}_{j}$ would mean that perturbations in $x_j$ will be able to affect a large portion of the system.\\

Note that in case of comparisons with other datasets, $\mathcal{D}_{j \rightarrow k}$, $\mathcal{D}^{*}_{j \rightarrow k}$ can be normalized by $1/\tau_{\infty}$; $\mathcal{D}_{j}$ can be normalized by the number of variables as $1/(N-1)$. 
These steps are not needed in this study.\\ 

\paragraph*{Link and strength maps.} Finally, for a given pattern $j$ identified by the dimensionality reduction strategy proposed in section \ref{sec:dim_red}, it is possible to plot the cumulative causal links $\mathcal{D}_{j \rightarrow k}$ and $\mathcal{D}^{*}_{j \rightarrow k}$ (Eq. \ref{eq:causal_strength_j_to_k_Kubo} and \ref{eq:causal_strength_j_to_k}) with any other pattern $k$ as a map. Given a pattern $j$ we will often refer to such map as ``link map'' $\mathcal{D}_{j \rightarrow k}$ for simplicity. Similarly, the ``causal strength'' $\mathcal{D}_{j}$ of each node $j$ as defined in Eq. \ref{eq:total_causal_strength_j} can be plotted as a map, referred to as ``strength map''.

\section{Data} \label{sec:data}

To explore and showcase the proposed causal framework we consider a long, stationary integration of the state-of-the-art coupled climate model GFDL-CM4 \cite{cm4}. The ocean component of CM4, named MOM6, has an horizontal grid spacing of 0.25$^\circ$ and 75 vertical layers \cite{om4}. The atmospheric/land component is the AM4 model \cite{AM4a,AM4b} with horizontal grid spacing of roughly 1$^\circ$ and 33 vertical layers. We consider the sea surface temperature field (SST) at global scale. The simulation considered, referred to as ``piControl'', is a 650 years long, stationary integration with constant $\text{CO}_2$ forcing set to preindustrial level. In this work we consider the last 300 years of this simulation. Even with stationary $\text{CO}_2$ forcing, the climate system can display variability at a vast range of time scales coming from the internal dynamics of the system. Importantly, especially at higher latitudes the system can display significant oscillations up to 10–100 years time scales, i.e. ``multidecadal oscillations'' \cite{multidecadal}. Even in a 300 years long run such low frequency oscillations are sampled only a few times. Therefore, to simplify the interpretation of results, in this work we high-pass filter every time series with a cut-off frequency of $f = 1/(10\text{ years})$ and focus on interannual variability only. Importantly, as shown in Appendix D, the distributions obtained after high-pass filtering each time series are well approximated by Gaussians, justifying the methodology proposed in section \ref{sec:quasiGaussian}. Furthermore, the analysis will focus on SST anomalies after removing the seasonal cycle (i.e., subtracting to each month its climatology). In this study we consider temporal resolution of 1 month as a reasonable time scale to observe propagation of signals among modes of variability at global scale.


\section{Causality in climate fields} \label{sec:climate}

\subsection{Applicability of fluctuation-response theory in climate studies} \label{sec:justification}

The main theoretical ideas justifying the application of methods in section \ref{sec:FDR} in climate, trace back at least to the work of Hasselmann, K. (1976) \cite{Hasselmann}. The main intuition of the ``Hasselmann's program'' \cite{LucariniChekroun} relies on thinking of processes with enough time scale separation between short and long time scales in terms of Brownian motion. Frankignoul and Hasselmann (1977) \cite{Hasselmann2} first showed that the statistical properties of sea surface temperature (SST) variability can be in fact explained (at first order) by linear stochastic models with white noise representing the fast atmospheric variability. Such ideas were further explored and convincingly demonstrated by Penland, C. (1989) \cite{Penland89} and Penland and Sardeshmukh (1995)  \cite{Penland95} and motivated recent work on coupling functions as in \cite{Wettlaufer1} and \cite{Keyes}.\\

The aforementioned studies justify the application of concepts introduced in section \ref{sec:FDR} to explore causality in climate fields. Specifically, this work will focus on the SST fields. Physically, this means that we will make the (rather strong) simplification of considering SST variability as a deterministic process and treat higher-frequency phenomena (e.g., atmospheric variability) as noise as done in \cite{Hasselmann}. Focusing only on sea surface temperature is however a limitation of this work and should be taken into account when analyzing the results. The extension to a multivariate framework is left for future work.

\subsection{Relation to previous climate studies} \label{sec:relation}

We briefly present the main relationship between fluctuation-dissipation response studies investigated in the climate literature \cite{Leith,Majda2010,MajdaBook,GRITSUN,Pedram2} and the causality framework explored here. Climate studies focused on studying the response $\delta \langle \boldsymbol{x}(t) \rangle$ of a dynamical system $\boldsymbol{x}$ perturbed by a small time-dependent forcing $\boldsymbol{f}$ as follows:
\begin{linenomath}
\begin{equation}
\delta \langle \boldsymbol{x}(t) \rangle = \int_{0}^{t} d\tau ~ \boldsymbol{R}(\tau) \boldsymbol{f}(t-\tau).
\label{eq:linear_response_climate}
\end{equation}
\end{linenomath}
Where $\boldsymbol{R}(t)$ is the linear response operator. In this study we consider stationary fields and \textit{impulse} perturbations and therefore the forcing can be written as a delta function $\delta(t-\tau)$. In such case, Eq. \ref{eq:linear_response_climate} reduces to:
\begin{linenomath}
\begin{equation}
\delta \langle \boldsymbol{x}(t) \rangle = \int_{0}^{t} d\tau ~ \boldsymbol{R}(\tau) \delta(t-\tau) = \boldsymbol{R}(t),
\label{eq:linear_response_climate_delta}
\end{equation}
\end{linenomath}
and the operator $\boldsymbol{R}(t)$ alone allows us to study causal links.\\

In what follows, responses in Eq. \ref{eq:linear_response_climate_delta} are computed by (a) using the quasi-Gaussian approximation as shown in Eq. \ref{eq:response_linear_markov} and (b) by first standardizing every time series to zero mean and unit variance. The operator $\boldsymbol{R}(t)$ itself will be non-dimensional.

\subsection{Application to global sea surface temperature} \label{sec:global}

\subsubsection{Dimensionality reduction and causal inference} \label{sec:dimensionality_red_global}

\begin{figure}[tbhp]
\centering
\includegraphics[width=1\linewidth]{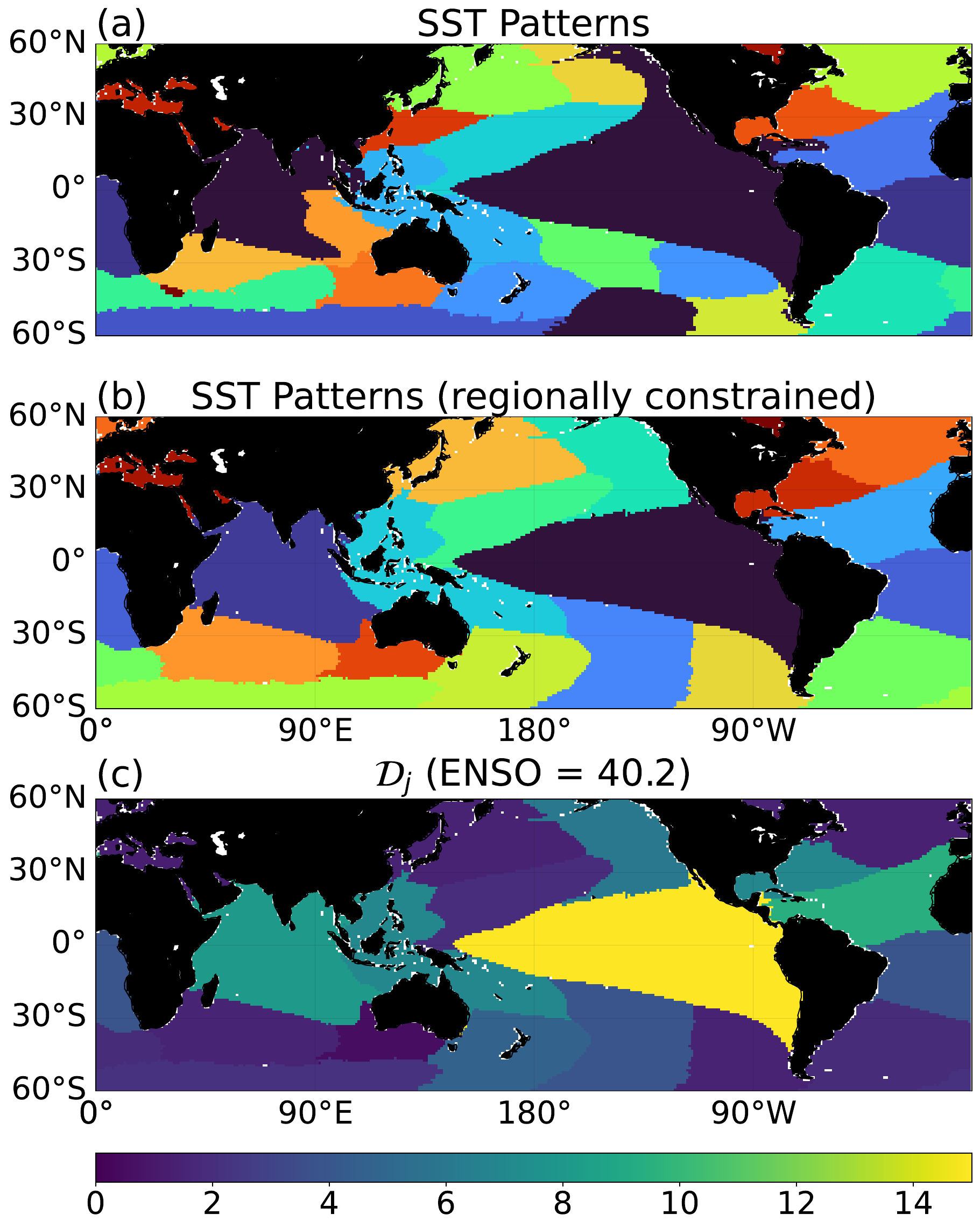}
\caption{Sea surface temperature (SST) patterns in the latitude range [60$^o$S-60$^o$N] and at monthly temporal resolution. Panel (a): an undirected graph is inferred through Eq. \ref{eq:adjacency_matrix} but without the proposed constraint $d(i,j)\leq \eta$. Then the community detection method Infomap is applied; see \ref{sec:dim_red}. Panel (b): same as panel (a) but the undirected graph is inferred through the newly proposed Eq. \ref{eq:adjacency_matrix}. Panel (c): causal strengths as defined by Eq. \ref{eq:total_causal_strength_j}. As expected the ``ENSO'' region is the strongest mode in the inferred causal network. Its strength is reported in the plot title. The response functions are computed up to $\tau_{\infty} = 10$ years. Only the statistical significant responses contribute to the strength metrics shown in Eq. \ref{eq:total_causal_strength_j}. Confidence bounds are quantified through Eq. \ref{eq:expectation_and_Variance_R} at the $\pm 3 \sigma$ level.}
\label{fig:global_communities_strengths}
\end{figure}

We now focus on sea surface temperature (SST) variability at global scale. We consider the latitudinal range 60$^o$S-60$^o$N at $1^o$ resolution accounting for $N = 31141$ time series. The SST field is saved as monthly averages for 300 years for a total of $T = 3612$ time steps. As a first step we aim in reducing the dimensionality of the field from $N = 31141$ to fewer components. First, we apply the community detection algorithm without constraining for the identification of regionally constrained patterns; in other words, the graph in Eq. \ref{eq:adjacency_matrix} is inferred solely by the correlations  between each time series $x_i(t)$ and $x_j(t)$ (i.e., term $C_{i,j} \geq k$ in Eq. \ref{eq:adjacency_matrix}). Applying the community detection algorithm without constraining on distances as proposed in this work results in patterns that are not regionally constrained as shown in Fig. \ref{fig:global_communities_strengths}. Fig. \ref{fig:global_communities_strengths}(a), shows the Indian Ocean, eastern Pacific and part of the Southern Ocean as part of the same pattern. The regional variability of such distant regions is indeed linked by physical processes (i.e., ``teleconnection'' patterns); for example at interannual time scales, Indian Ocean variability is forced by the tropical Pacific through an atmospheric wave response to El Ni\~no events \cite{Chiang}. Consequently, the sea surface temperature variability in such regions is often grouped under the same cluster by dimensionality reduction algorithms. The constraint on distances proposed here, $d(i,j)\leq \eta$ in Eq. \ref{eq:adjacency_matrix}, allows us for the identification of local and spatially contiguous patterns as shown in Fig.  \ref{fig:global_communities_strengths}(b), so that the Indian Ocean, eastern Pacific and part of the Southern Ocean are now all captured as different clusters. Therefore, the additional constraint $d(i,j)\leq \eta$ introduced in Eq. \ref{eq:adjacency_matrix} is a simple but important step when coarse graining the system. The proposed dimensionality reduction method allows us to reduce the dimensionality from $N = 31141$ to $N = 19$ time series. The patterns identified are \textit{regionally constrained}, therefore allowing us to answer the following question: how does the climate system respond to \textit{local} perturbations? To answer such question, we leverage the tools presented in section \ref{sec:FDR}.\\

We consider the fluctuation-dissipation relation in its quasi-Gaussian approximation as shown in Eq. \ref{eq:response_linear_markov}. In the Appendix, section \ref{app:histograms_global} we show that the time series of each pattern (i.e., mode) follows approximately a Gaussian distribution, therefore justifying the quasi-Gaussian approximation. We infer causality up to a $\tau_{\infty} = 10$ years and show the causal strength $\mathcal{D}_{j}$ (Eq. \ref{eq:total_causal_strength_j}) in Fig. \ref{fig:global_communities_strengths}(c). The strongest mode of variability at interannual time scales is in the tropical Pacific, as expected \cite{ENSOcomplexity}. Physically, results in Fig.\ref{fig:global_communities_strengths}(c) imply that the variability in the tropical Pacific is able to influence a larger part of the world compared to other regions with smaller strength. In what follows we are going to refer to this region as ``ENSO region''.\\

\subsubsection{Investigation of causal interactions} \label{sec:inidividual_links}

We further analyze the links between three components of the system. Specifically, we focus on the interaction of ENSO, the Indian Ocean (IO) and South Tropical Atlantic (STA). ENSO is known to drive climate variability outside the tropical Pacific through teleconnection patterns and has been studied in many contributions. The way in which Indian and Atlantic variability drive SST in the Pacific has been less appreciated in the past and it is currently debated in the community \cite{Pantropical}. Quantification of such linkages is important to better understand climate variability and to improve seasonal forecasting.\\

\begin{figure*}[tbhp]
\centering
\includegraphics[width=1\linewidth]{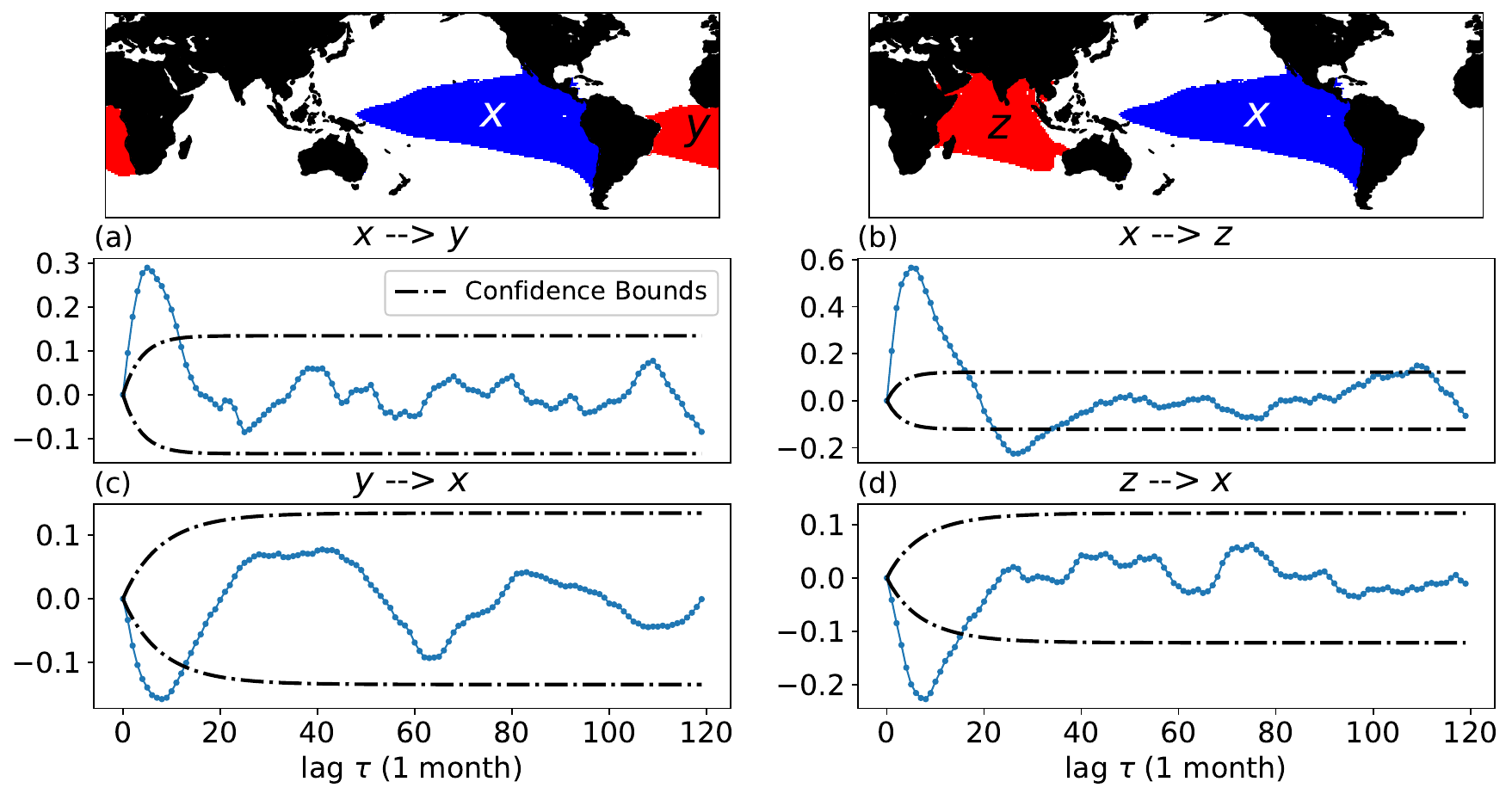}
\caption{$x$: ENSO mode. $y$: South Tropical Atlantic. $z$: Indian Ocean. Panel (a,c): causal link $x \rightarrow y$ and $y \rightarrow x$. Panel (b,d): causal link $x \rightarrow z$ and $z \rightarrow x$. Response functions have been computed up until $\tau_{\infty} = 10$ years. Confidence bounds are quantified through Eq. \ref{eq:expectation_and_Variance_R} at the $\pm 3 \sigma$ level. Responses in between the confidence bounds are here considered as spurious.}
\label{fig:ENSO_links}
\end{figure*}

During an El Ni\~no phase, the anomalous temperature in the tropical Pacific excites waves in the atmosphere. Such waves, known as eastward-propagating Kelvin and westward-propagating Rossby waves, drive changes in temperature in the whole tropical band \cite{Chiang}. Such causal links are identified in Fig. \ref{fig:ENSO_links}(a,b), with positive responses of both the IO and STA regions to perturbations in the ENSO regions. As expected such positive lead of ENSO is the strongest in magnitude and much larger than the other responses in Fig. \ref{fig:ENSO_links}. Interestingly, we find a (weak) negative link between ENSO and IO in Fig. \ref{fig:ENSO_links}(b) around $\tau = 30$ months, suggesting the emergence of positive anomalies in the Indian Ocean $\sim 3$ years after La Ni\~na events, and viceversa for El Ni\~no events. The positive response around 10 years in Fig. \ref{fig:ENSO_links}(b) is considered as a False Positive.\\ 

Fig. \ref{fig:ENSO_links}(c) shows that the anomalies in the STA region, mainly linked to the dynamics of the Atlantic Ni\~no \cite{AtlanticNino} (see also discussion in \cite{Bracco}), lead \textit{on average} to the development of anomalies of the opposite sign in the eastern Pacific as recently argued in the literature \cite{Fonseca,Keenlyside,Ham}.\\

The IO pattern in our study (see pattern $z$ in Figure \ref{fig:ENSO_links}) mainly identifies what is known as the Indian Ocean Basin (IOB) mode \cite{Klein}. The IOB mode has been traditionally considered as simply forced by ENSO. Nonetheless, recent studies have revealed how the IOB can also drive ENSO variability. Specifically, it has been demonstrated how a strong IOB warming can in fact contribute to central Pacific cooling further driving a transition to a La Ni\~na state \cite{Izumo,Ha,Pantropical}. Such negative link is correctly identified by the proposed framework (see Fig. \ref{fig:ENSO_links}(d)) but does not show up in correlation-only analyses (see for example Fig. 11(b) in \cite{dimRedClimate}). As discussed also in \cite{Pantropical} these results suggest an increase in potential predictability of ENSO variability when considering the non-local interactions with the Indian Ocean and tropical Atlantic basins. We note that the results shown here come from a climate model which, as any models, it is far from perfect. The causal links shown in Fig. \ref{fig:ENSO_links} agree with the existing literature in terms of directionality of the links and sign, however the leading of the tropical Pacific to the Indian and Atlantic basins in this model may be overestimated as shown in another climate model (by correlation analysis-only) in Falasca et al. (2019) \cite{dimRedClimate}. Future work may focus on the proposed framework to compare model and reanalyses data, similar to what was done in Falasca et al. (2019) \cite{dimRedClimate}.\\

\begin{figure*}[tbhp]
\centering
\includegraphics[width=1\linewidth]{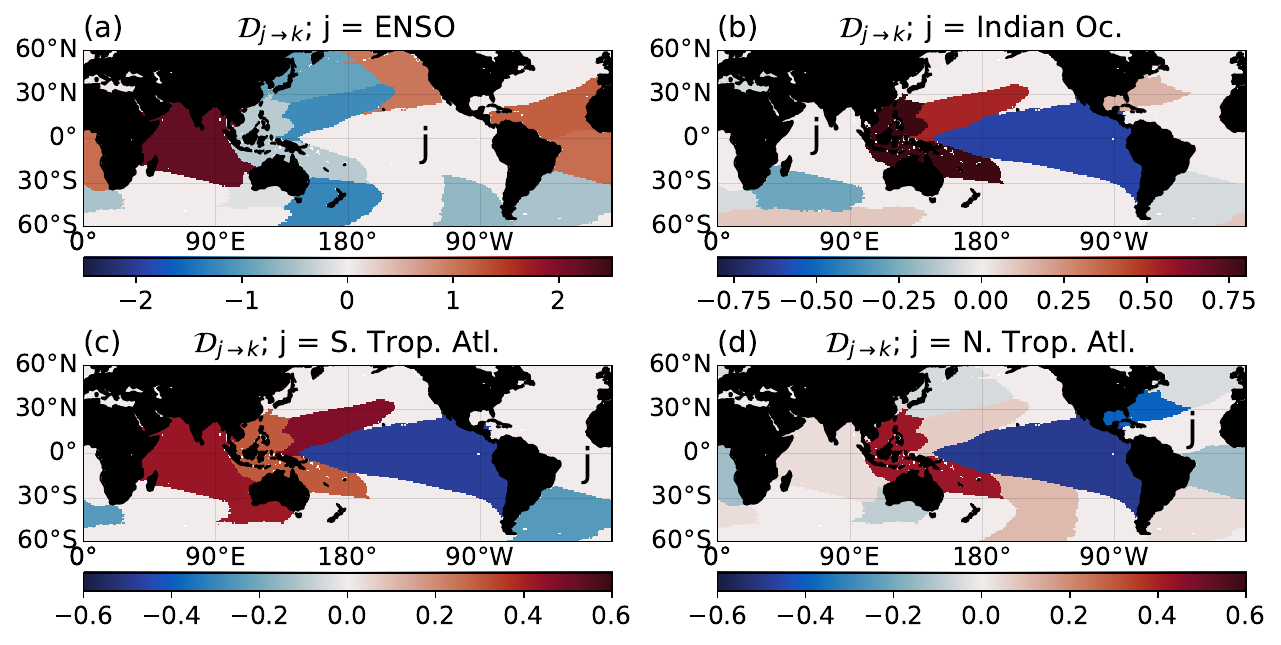}
\caption{Link maps $\mathcal{D}_{j \rightarrow k}$ for all $k$, as computed in \ref{eq:causal_strength_j_to_k_Kubo} and considering only up to $\tau_{\infty} = 6$ months. Regions $j$ considered are ENSO region, Indian Ocean, South and North Tropical Atlantic in panels (a,b,c,d) respectively. The first Empirical Orthogonal Function roughly correspond to the ENSO link map in panel (a). Only the statistical significant responses contribute to the causal link maps. Confidence bounds are quantified through Eq. \ref{eq:expectation_and_Variance_R} at the $\pm 3 \sigma$ level.}
\label{fig:causal_link_map_6months}
\end{figure*}

Finally, in Fig. \ref{fig:causal_link_map_6months} we show the cumulative response of the whole system to the climate variability in four regions:  ENSO region, Indian Ocean (IO), South and North Tropical Atlantic (STA and NTA respectively). Such ``link maps'', introduced at the end of section \ref{sec:metrics},  allow us to visualize the cumulative degree of causation $\mathcal{D}_{j \rightarrow k}$ (Eq. \ref{eq:causal_strength_j_to_k_Kubo}) up to a time lag $\tau_{\infty}$, here chosen as $\tau_{\infty} = 6$ months. Fig. \ref{fig:causal_link_map_6months}(a) quantifies the cumulative response of any region given perturbations in the ENSO region. We notice that such map is qualitatively similar to the first Empirical Orthogonal Function of global SST (see for example Fig. 4 in \cite{SSTEOF}). The framework allows us to examine causal linkages from/to any region of the system. Figures \ref{fig:causal_link_map_6months}(b,c,d) show the cumulative degree of causation respectively from  IO, STA and NTA regions to any other region in the world. In other words, such link maps allow us to summarize the cumulative response of the whole system, given small, local perturbations to any region $x_j$ of choice, offering a useful and simplified approach to explore climate dynamics from data.

\section{Conclusions and discussion} \label{sec:conclusions}

This work introduces a novel framework for causal inference in spatiotemporal climate fields. The method relies on two independent steps: dimensionality reduction and causal inference. The causal inference step, based upon ideas of Baldovin et al. \cite{Baldovin} frames the problem of causality in the formalism of linear response theory \cite{KuboBook}. Here, we further developed these ideas by proposing an analytical \textit{null} model for the fluctuation-dissipation relation. The model allows us to distinguish between true and spurious response functions estimated from finite data, with applicability not restricted to climate. Causality is inferred after reducing the dimensionality of the system into a few \textit{regional patterns}, i.e., proxies of ``modes'' of variability. Such ``modes'' are defined as regionally constrained sets of time series with large average pairwise correlation. The  dimensionality reduction and the causal inference steps allow to study how \textit{local} perturbations can propagate through the system and impact \textit{remote} locations.\\

We discuss a few important limitations and caveats that may hinder interpretations of results in future studies.\\

\paragraph{The case of hidden variables.} The fluctuation-dissipation formalism identifies causal links when we have access to the whole state vector $\boldsymbol{x}$. However, often in real-world cases we can access only a few variables. A solution is to include the proper variables for the phenomena we want to explain \cite{Baldovin}. In this work, we based our analysis on sea surface temperature building on ideas first proposed by Hasselmann, K. (1977) \cite{Hasselmann2,LucariniChekroun} where the fast atmospheric variability can be considered as noise, forcing the slower deterministic ocean dynamics. Therefore, given the sea surface temperature field, the size of the spatial patterns considered and their temporal resolution (i.e., $\Delta T = 1$ month) we assume the system to be approximately Markovian. However, the focus on sea surface temperature only is a great simplification and should be considered when interpreting results. The question of how many variables are enough to consider the system as Markovian is an old problem with warnings discussed at least since Onsager and Machlup (1953) \cite{Onsager}; see also section IVB in \cite{Baldovin}. Quite interestingly, \cite{Baldovin} also showed that applying Takens theorem \cite{Takens} to reconstruct the state space vector may not always help. The main reason is that Takens embedding theorem, proven for deterministic systems \cite{Takens}, fails for general stochastic processes \cite{Baldovin}. More recent versions of Takens theorem have been proved for stochastic systems and could be potentially explored in future studies, see for example \cite{Robinson,Stark}\\

\paragraph{Computation of the inverse covariance matrix $\boldsymbol{C}(0)^{-1}$.} Consider a dynamical system $\boldsymbol{x} \in \mathbb{R}^{N,T}$, $N$ is its dimensionality and $T$ is the length of each time series $x_{i}(t)$. If $N>T$, the covariance matrix $\boldsymbol{C}(0) \in \mathbb{R}^{N,N}$ will not be full rank, and therefore it will not have an inverse. Generally, the covariance matrix can be ill-conditioned and the computation of the inverse $\boldsymbol{C}(0)^{-1}$ will result in large errors \cite{GRITSUN,Pedram2,Martynov1,Martynov2}. Therefore, the proposed framework should be applied for systems $\boldsymbol{x} \in \mathbb{R}^{N,T}$ with $T\gg N$, i.e., the number of samples much larger than the dimensionality of the system. As a simple test, when computing responses with the quasi-Gaussian approximation $\boldsymbol{R}(\tau) = \boldsymbol{C}(\tau)\boldsymbol{C}(0)^{-1}$ we recommend to check $\boldsymbol{R}(0) = \boldsymbol{I}$ (at least up to a certain numerical accuracy), $\boldsymbol{I}$ being the Identity matrix. In general, dimensionality reduction schemes (as proposed in this paper) reduce the number of time series $N$ to values much smaller than $T$, allowing for trustworthy computations of $\boldsymbol{C}(0)^{-1}$ \cite{GRITSUN}.\\

\paragraph{Quasi-Gaussian approximation.} The fluctuation-dissipation relation used in this study is valid for linear systems and it has been shown to work well also for weakly nonlinear systems, see discussion in section \ref{sec:quasiGaussian}. However, before applying the methodology proposed here we suggest careful analysis of the data distribution to avoid misleading results. An example is the work shown in \cite{Provenz}, where the authors analyzed the causal link between $\text{CO}_2$, temperature ($T$) and insolation in the last $800$ kyr using the Fluctuation-Response formalism \cite{Baldovin}. Distributions of both $\text{CO}_2$ and $T$ in the last $800$ kyr are strongly non-Gaussian. The solution was to high-pass filter the data and focus on high-frequency variability, with the hypothesis of slow time scales being linked to the external forcing and faster time scales to the internal variability of the system. The temporal filtering was shown to be enough to recover Gaussian distributions \cite{Provenz}. In this work, we also high-pass filtered the data with a cut-off frequency of $f = 1/(10\text{ years})$. The probability distributions obtained after the filtering can be reasonably approximated by Gaussians (see Appendix \ref{app:histograms_global}), justifying the application of the methodology shown in section \ref{sec:FDR}. A generalization to nonlinear systems is provided by formula \ref{eq:response_general}, as long as the probability distribution $\rho(\boldsymbol{x})$ is known. In specific cases, we note that it is possible to apply transformations to strongly non-Gaussian fields and still use the quasi-Gaussian approximation explored here. An example is the precipitation field, where a logarithmic scaling can help recover Gaussian-like distributions \cite{Pendergrass}.\\

The methodology proposed here can be applied to study the dynamics of any climate field, at least given the assumptions and limitations listed above. It serves as a useful, rigorous framework to simplify the description of complex, high-dimensional dynamical systems in terms of a few entities and their linkages, aiming to better understand the system's dynamics. Unlike other methods for causal discovery adopted in climate, the proposed scheme scales to high-dimensional datasets; in fact Fluctuation-Dissipation formulas have been shown to scale to thousands of time series in climate applications (e.g., \cite{GRITSUN}). Moreover, the causal inference method and the proposed \textit{null} model have a clear physical interpretation, they are formalized via analytical formulas and they can be easily implemented without the need for many heuristics and parameters.\\ 

The application explored in section \ref{sec:global} allowed us to detect well-known links in climate, such as the influence of tropical Pacific variability onto other basins, as well as other linkages, such as the lead of sea surface temperature variability in the Indian Ocean to the Pacific basin, which received less attention in the literature \cite{Pantropical}. Additionally, we showed how the ``strength maps'' and ``link maps'' as shown in Fig. \ref{fig:global_communities_strengths}(c) and Fig. \ref{fig:causal_link_map_6months} summarize cumulative causal interactions across time and space in a comprehensive and interpretable way.\\

We focused on the sea surface temperature field as the statistics of modes of variability and their linkages in this field have been investigated in many previous studies, therefore offering a good test-case for the methodology. Importantly, climate studies often focus on a few modes at a time (e.g. \cite{Pantropical} and references therein). Here we showed that the methodology allows us to study causal linkages among regions in a comprehensive framework, where all modes of variability and their interactions are studied simultaneously.\\

Examples of future work range from studying the evolution of climate modes and their linkages in paleoclimate simulations, with time-dependent orbital and trace-gases forcings (e.g., \cite{Falasca}), to replacing expensive Green’s function approaches to diagnose relationships among variables and their sensitivity to external forcings \cite{Ming}. Finally, the proposed framework also offers a way to evaluate new generations of climate models in terms of their emergent causal structure; for example, by assessing the impact of new sub-grid parametrizations onto the large scale dynamics.

\begin{acknowledgments}
FF acknowledges helpful discussions with Marco Baldovin, Simone Contu, Andre Souza, Pedram Hassanzadeh, Aurora Basinski and Chris Pedersen. FF also thanks Martin Rosvall for clarifications on the Infomap methodology at the start of this work. This work was supported in part by NOAA grant NOAA-OAR-CPO-2019-2005530, by the KITP Program “Machine Learning and the Physics of Climate” supported by the National
Science Foundation under Grant No. NSF PHY-1748958 and by Schmidt Futures, a philanthropic initiative founded by Eric and Wendy Schmidt, as part of its Virtual Earth System Research Institute (VESRI). Finally, we thank two anonymous
reviewers for their insightful suggestions and comments.

\end{acknowledgments}

\section*{Code availability}
Codes and materials are available at \href{https://github.com/FabriFalasca/Linear-Response-and-Causal-Inference}{https://github.com/FabriFalasca/Linear-Response-and-Causal-Inference}.

\appendix

\section{Dimensionality reduction in climate. Limitations of current methods and proposal} \label{app:dim_red_methods}

\subsection{Two goals in dimensionality reduction studies} \label{sec:dim_red_goals}
We note that the use of dimensionality reduction in applications of linear response theory can be leveraged with at least two different goals in mind. In the case of very high dimensional systems as a General Circulation Model (GCM), applications of the fluctuation-dissipation response formalism is practically impossible. The usual solution in the climate literature has been to construct response operators in a low-dimensional space spanned by many Empirical Orthogonal Functions (EOFs or Principal Components) \cite{PCA}; usually, order $10^3$ EOFs in order to explain at least 90$\%$ of the total variance. Results computed in the low dimensional space are then transformed back to the original space \cite{GRITSUN, Majda2010, Pedram2}. This computational strategy has been shown to be successful in many applications (see \cite{GRITSUN,Majda2010}). A second possible goal of dimensionality reduction is to simplify the problem in hand in terms of very few components and apply the linear response formalism directly on those entities. In this case we are interested in studying directly the coarse-grained version of the system. This adds to interpretability and to a first order understanding of the system's dynamics. This second case is the one considered in this paper.
\subsection{Few limitations of common methods and proposal} 
Traditionally, dimensionality reduction in climate studies is done through Principal Component Analysis (PCA) \cite{PCA}. PCA, or Empirical Orthogonal Function (EOF) analysis \cite{storchzwiers} is a useful, first order way to reduce the dimensionality of the system based on the singular value decomposition (see e.g., \cite{Witten}) of the data matrix. However, the resulting patterns suffer from a few drawbacks: first, EOFs are orthogonal by definition. Such constraint hamper their interpretation and make it difficult to distinguish between physical or purely statistical modes \cite{Dommenget,Tantet}. A possible solution has been to rotate the EOFs, such as in \cite{Kawamura}. Rotated-EOFs have been found to be sensitive to the rotation criterion, normalizations and number of loadings (see \cite{storchzwiers,Tantet}).\\ 

Another drawback comes from linearity. Manifold learning algorithms aim in addressing this issue by identifying low-dimensional representations of a high-dimensional system accounting for nonlinearities (curved manifolds) \cite{Saul}. Examples range from the Isomap algorithm \cite{Tenenbaum} to the more recent t-SNE \cite{tSNE}, UMAP \cite{UMAP} to the PHATE algorithm \cite{PHATE} and ROCK-PCA \cite{ROCK-PCA}. Finally, deep learning tools such as autoencoders can be explored for dimensionality reduction \cite{AE} and found applications in climate science \cite{sara}.\\

Dependent on the goal in mind (see section \ref{sec:dim_red_goals}), a possible limitation shared by all these tools when applied to global climate data is that they decompose a field in terms of \textit{global} (in longitude-latitude maps) modes. However, physically, climate dynamics can be often thought of as a set of \textit{remote} connections driven by \textit{local} phenomena (perturbations). Given so, common practice in climate science has been to define ``climate indices'' as time series averaged in specified regions (i.e., ``boxes''). Known examples are the Ni\~no3.4, the Indian Ocean Dipole (IOD) index etc. However, a framework for automated identification of proxies of such indices is needed as the locations of such regions, or ``boxes'', may be not relevant for the study of future (or past) climates. An example can be found in \cite{Thirumalai,DiNezio,Falasca} where the authors showed the emergence of an El Ni\~no-like variability in the Indian Ocean during the Last Glacial Maximum, the last 6000 years and in future projections. In this sense, known indices identified in the current climate are potentially less meaningful in past and future climates.\\

A method proposed to automatically identify proxies for climate indices is $\delta$-MAPS \cite{Ilias_1}. Given a climate fields, $\delta$-MAPS identifies spatially contiguous clusters. The method has proven to be useful in climate studies with applications ranging from model evaluation \cite{dimRedClimate,Dalelane}, shifts in climate modes in the last 6000 years \cite{FalascaEPJ,Falasca}, sea level budget at regional scale \cite{Camargo},  marine ecology \cite{ljuba1} and ecosystem dynamics \cite{ljuba2}. In the case of relatively low dimensional fields (e.g., global fields at $2^\circ$ by $2^\circ$ spatial resolution) $\delta$-MAPS shows excellent performance. However, a known drawback is that it does not scale well with high-dimensional datasets (i.e., large number of grid cells). Additionally, exploratory tests are needed to explore the sensitivity to parameter choices in the domain identification step.\\

When working with very high dimensional fields, it is often useful to consider fast and scalable algorithms. In the last two decades, climate data analysis have focused on fast methodologies stemming from the complex network literature \cite{Barabasi}. An example is the work of \cite{Tantet} where the authors focused on the community detection method ``Infomap'' \cite{Rosvall1,Rosvall2,mapequation2022software} to identify communities in the HadISST \cite{HadISST} sea surface temperature dataset. Such methods allow us to find patterns that are not necessarily orthogonal. Furthermore, they are fast, memory efficient and scale well with the dimensionality of the dataset. The main issue is that, similar to manifold learning algorithms, community detection algorithms are not constrained to be spatially contiguous \cite{Ilias_1}.\\

In this paper we showed that adding a simple constraint on spatial distances is enough to enforce the identification of ``local'' patterns (see section \ref{sec:graph_inference}). This allows us to leverage computationally fast and robust methods such as community detection for dimensionality reduction strategies in climate. Differently from $\delta$-MAPS \cite{Ilias_1}, the identified patterns cannot overlap with each other. We find however that conclusions found in previous studies using $\delta$-MAPS (see \cite{dimRedClimate} for example) may not be strongly dependent on clustering overlapping, at least when focusing on the sea surface temperature field. The framework proposed here in section \ref{sec:dim_red} is then leveraged as a much simpler (and therefore more robust), practical framework to the problem of identification of \textit{regionally constrained} modes.

\section{A \textit{null} model for the Fluctuation-Dissipation relation. Analytical derivation of the confidence bounds} \label{app:confidence_bounds}

This work proposes a novel null model for the Fluctuation-Dissipation relation (see \ref{eq:linear_markov_matrix_red_noise}). In the null model, every variable $x_j$ and $x_k$ is independent, and therefore the expected value of each response $\mathbb{E}[R_{k,j}(\tau)] = 0$ for $j \neq k$ by construction. Nonetheless, estimating such responses by $\boldsymbol{R}(\tau) = \boldsymbol{C}(\tau)\boldsymbol{C}(0)^{-1}$ (see \ref{sec:quasiGaussian}) using time series of finite length $T$ simulated by the null model, will give rise to spurious results diverging from the expected value $\mathbb{E}[R_{k,j}(\tau)]$. In Eq. \ref{eq:expectation_and_Variance_R} of the main text we showed the analytical probability distribution of $R_{k,j}(\tau)$. The main assumption in this derivation is that responses $R_{k,j}(\tau)$ follow a Normal distribution. Therefore the expected value $\mathbb{E}[R_{k,j}(\tau)]$ and variance $\mathbb{V}\textrm{ar}[R_{k,j}(\tau)]$ uniquely define the probability density $\rho(R_{k,j}(\tau))$. Here we present the derivation of such formula.
\subsection{Notation adopted in this section} \label{app:notation}

In order to simplify and ease the derivation, it is useful to adopt a simpler and more appropriate statistical formalism. The symbols adopted in this section relate to the ones used in the previous ones as follows: $\mathbb{E}[X] = \langle X \rangle$ represents the expected value of a random variable $X$. This is equal to the ensemble average considered in the previous sections. Consequently, $\mathbb{V}\textrm{ar}[X] = \mathbb{E}[(X - \mathbb{E}[X])^{2}]$ represents the variance of a random variable $X$. Finally, $\mathbb{C}\textrm{ov}[X,Y] = \mathbb{E}[(X - \mathbb{E}[X]) (Y - \mathbb{E}[Y])]$ represents the covariance of two random variables $X$ and $Y$. We are going to refer to the null process as $\boldsymbol{x} = [x_{1}(t),x_{2}(t),...,x_{N}(t)]$ (rather than $\boldsymbol{\Tilde{x}}$ as in \ref{eq:linear_markov_matrix_red_noise}). Finally, each time series $x_{j}(t)$ is here considered to be scaled to zero mean and unit variance. This step greatly simplifies the derivation. At the end of this section, we provide the general formula for processes that are not unit-variance.

\subsection{Analytical derivation} \label{app:derivation_bounds}

Consider a long trajectory $\boldsymbol{x} \in \mathbb{R}^{N,T}$ defined by the forward iteration of the \textit{null} model in Eq. \ref{eq:linear_markov_matrix_red_noise}. The \textit{true} mean, and covariances at lag $\tau$ of each individual time series in $\boldsymbol{x}$ are given by $\mathbb{E}[x_{j}(t)] = 0$ and $\mathbb{E}[x_{k}(t+\tau) x_{j}(t)] =  \phi_{k}^{\tau} \delta_{k,j} $ respectively. Where $\phi_{k}$ is the lag-1 autocorrelation of time series $x_{k}(t)$ and the Kronecker delta $\delta_{k,j}$ differs from zero only in the case $j = k$.\\

We note that the numerical estimation of both $\boldsymbol{C}(\tau)$ and $\boldsymbol{C}(0)^{-1}$ will lead to spurious terms in $\boldsymbol{R}(\tau)$. We then rewrite the covariance matrix $\boldsymbol{C}(\tau)$ estimated through time averages as a sum of the expected value $\mathbb{E}[\boldsymbol{C}(\tau)]$ plus some small Gaussian residual $\boldsymbol{\hat{C}}(\tau)$ as:
\begin{linenomath}
\begin{equation}
\boldsymbol{C}(\tau) = \mathbb{E}[\boldsymbol{C}(\tau)] + \boldsymbol{\hat{C}}(\tau) = \boldsymbol{D}_{\phi}^{\tau} + \boldsymbol{\hat{C}}(\tau).
\label{eq:covariance_plus_residuals} 
\end{equation}
\end{linenomath}
Where $\boldsymbol{D}_{\phi}^{\tau}$ is a diagonal matrix with component $(i,j)$ defined as $(\boldsymbol{D}_{\phi}^{\tau})_{i,j} = \phi_{i}^{\tau}\delta_{i,j}$. The decomposition (Eq. \eqref{eq:covariance_plus_residuals}) applies to the matrix $\boldsymbol{C}(0)$ as well with $\boldsymbol{D}_{\phi}^{0} = \boldsymbol{I}$ where $\boldsymbol{I}$ is the Identity matrix. The main difficulty is that we are not interested in $\boldsymbol{C}(0)$ but in its inverse $\boldsymbol{C}(0)^{-1}$. By assuming relatively small residuals (true for time series with $T \gg 1$), we can approximate an inverse of the estimated covariance matrix $\boldsymbol{C}(0)^{-1}$ using Neumann series \cite{methodsMathPhysics} as:
\begin{linenomath}
\begin{equation}
\boldsymbol{C}(0)^{-1} = (\boldsymbol{I} + \boldsymbol{\hat{C}}(0))^{-1} \approx \boldsymbol{I} - \boldsymbol{\hat{C}}(0).
\label{eq:inverse_covariance_lag0_plus_residuals} 
\end{equation}
\end{linenomath}
Where we only retained the first term in the Neumann series. An estimator of the null response $\boldsymbol{R}(\tau) = \boldsymbol{C}(\tau) \boldsymbol{C}(0)^{-1}$ can be then written as 
\begin{linenomath}
\begin{equation}
\boldsymbol{R}(\tau) = \boldsymbol{C}(\tau) \boldsymbol{C}(0)^{-1} \approx \boldsymbol{C}(\tau) + \boldsymbol{D}_{\phi}^{\tau} (\boldsymbol{I} - \boldsymbol{C}(0)).
\label{eq:estimator_R} 
\end{equation}
\end{linenomath}
Where we neglected the term $\boldsymbol{\hat{C}}(\tau) \boldsymbol{\hat{C}}(0)$, a reasonable step in the presence of small residuals, true for time series with length $T \gg 1$. To derive the statistical properties of the estimator in Eq. \ref{eq:estimator_R}, it is useful to rewrite such formula in terms of each component $j$ and $k$. 
\begin{linenomath}
\begin{equation}
R_{k,j}(\tau) \approx C_{k,j}(\tau) + \delta_{k,j}\phi_{k}^{\tau} - \phi_{k}^{\tau} C_{k,j}(0).
\label{eq:pointwise_estimator_R} 
\end{equation}
\end{linenomath}
The final step is to derive the expected value $\mathbb{E}[R_{k,j}(\tau)]$ and $\mathbb{V}\textrm{ar}[R_{k,j}(\tau)]$ of Eq. \ref{eq:pointwise_estimator_R}, thus uniquely defining the probability distribution of $R_{k,j}(\tau)$, under the assumption of Gaussian statistics.

\subsubsection{Expected value and variance of the response estimator} \label{app:null_probability}

The expectation of the response estimator proposed in \ref{eq:pointwise_estimator_R} can be derived as 
\begin{linenomath}    
\begin{equation}
\begin{split}
\mathbb{E}[R_{k,j}(\tau)] &= \mathbb{E}[C_{k,j}(\tau)] + \delta_{k,j}\phi_{k}^{\tau} - \phi_{k}^{\tau} \mathbb{E}[C_{k,j}(0)] \\
&= \delta_{k,j} \phi_{k}^{\tau} + \delta_{k,j}\phi_{k}^{\tau} - \phi_{k}^{\tau} \delta_{k,j} \\
&= \delta_{k,j} \phi_{k}^{\tau}.
\label{eq:pointwise_estimator_R_Expectation} 
\end{split}
\end{equation}
\end{linenomath}
The variance of the response estimator proposed in \ref{eq:pointwise_estimator_R} can be derived as 
\begin{linenomath}     
\begin{equation}
\begin{split}
\mathbb{V}\textrm{ar}[R_{k,j}(\tau)] &= \mathbb{V}\textrm{ar}[C_{k,j}(\tau) - \phi_{k}^{\tau} C_{k,j}(0)] \\
&= \mathbb{V}\textrm{ar}[C_{k,j}(\tau)] + \phi_{k}^{2\tau}\mathbb{V}\textrm{ar}[C_{k,j}(0)] \\ 
&- 2\phi_{k}^{\tau}\mathbb{C}\textrm{ov}[C_{k,j}(\tau),C_{k,j}(0)].
\label{eq:pointwise_estimator_R_Variance} 
\end{split}
\end{equation}
\end{linenomath}
We remind the reader the following useful equality: the covariance $\mathbb{C}\textrm{ov}[X,Y]$ of two random variables $X$ and $Y$ can be rewritten as $\mathbb{C}\textrm{ov}[X,Y] = \mathbb{E}[X Y] - \mathbb{E}[X]\mathbb{E}[Y]$. We now compute the variance of the response estimator in Eq. \ref{eq:pointwise_estimator_R_Variance}. To do so, we first need to provide an expression to terms $\mathbb{V}\textrm{ar}[C_{k,j}(\tau)]$ and $\mathbb{C}\textrm{ov}[C_{k,j}(\tau), C_{k,j}(0)]$. Such terms can be computed as follows:

\begin{widetext}
\begin{linenomath}
\begin{equation}
\begin{split}
 \mathbb{V}\textrm{ar}[C_{k,j}(\tau)] &= \mathbb{E}[C_{k,j}(\tau) C_{k,j}(\tau)] - \delta_{k,j} \phi_{k}^{2\tau} \\
 &= \frac{1}{T^2} \sum_{t',t'' = 1}^{T} \mathbb{E}[x_k(t'+\tau)x_j(t')x_k(t''+\tau)x_j(t'')] - \delta_{k,j} \phi_{k}^{2\tau} \\
 &= \frac{1}{T^2} \sum_{t',t'' = 1}^{T}\Big{(}\mathbb{E}[x_k(t'+\tau) x_k(t''+\tau)] \mathbb{E}[x_j(t')x_j(t'')] \\ 
 &+ \mathbb{E}[x_k(t'+\tau) x_j(t')] \mathbb{E}[x_k(t''+\tau)x_j(t'')] \\ 
 &+ \mathbb{E}[x_k(t'+\tau) x_j(t'')] \mathbb{E}[x_j(t')x_k(t''+\tau)]\Big{)} - \delta_{k,j} \phi_{k}^{2\tau} \\
 &= \frac{1}{T^2} \sum_{t',t'' = 1}^{T}\Big{(} \phi_k^{\mid t' - t'' \mid} \phi_j^{\mid t' - t'' \mid} + \delta_{k,j}\phi_k^{2\tau} + \delta_{k,j}\phi_k^{\mid t' +\tau - t'' \mid} \phi_j^{\mid t' -\tau - t'' \mid}\Big{)} - \delta_{k,j} \phi_{k}^{2\tau} \\
 &= \frac{1}{T^2} \sum_{t',t'' = 1}^{T}\Big{(} \phi_k^{\mid t' - t'' \mid} \phi_j^{\mid t' - t'' \mid} + \delta_{k,j}\phi_k^{\mid t' +\tau - t'' \mid} \phi_j^{\mid t' -\tau - t'' \mid}\Big{)}.
\label{eq:computaton_variance} 
\end{split}
\end{equation}
\end{linenomath}
\begin{linenomath}
\begin{equation}
\begin{split}
 \mathbb{C}\textrm{ov}[C_{k,j}(\tau), C_{k,j}(0)] &= \mathbb{E}[C_{k,j}(\tau) C_{k,j}(0)] - \delta_{k,j} \phi_{k}^{\tau} \\
 &= \frac{1}{T^2} \sum_{t',t'' = 1}^{T} \mathbb{E}[x_k(t'+\tau)x_j(t')x_k(t'')x_j(t'')] - \delta_{k,j} \phi_{k}^{\tau} \\
 &= \frac{1}{T^2} \sum_{t',t'' = 1}^{T}\Big{(}\mathbb{E}[x_k(t'+\tau) x_k(t'')] \mathbb{E}[x_j(t')x_j(t'')] \\ 
 &+ \mathbb{E}[x_k(t'+\tau) x_j(t')] \mathbb{E}[x_k(t'')x_j(t'')] \\ 
 &+ \mathbb{E}[x_k(t'+\tau) x_j(t'')] \mathbb{E}[x_j(t')x_k(t'')]\Big{)} - \delta_{k,j} \phi_{k}^{\tau} \\ 
 &= \frac{1}{T^2} \sum_{t',t'' = 1}^{T}\Big{(} \phi_k^{\mid t' +\tau - t'' \mid} \phi_j^{\mid t' - t'' \mid} + \delta_{k,j}\phi_k^{\tau} + \delta_{k,j}\phi_k^{\mid t' +\tau - t'' \mid} \phi_j^{\mid t' - t'' \mid}\Big{)} - \delta_{k,j} \phi_{k}^{\tau} \\
 &= \frac{1}{T^2} \sum_{t',t'' = 1}^{T}\Big{(} \phi_k^{\mid t' +\tau - t'' \mid} \phi_j^{\mid t' - t'' \mid} + \delta_{k,j}\phi_k^{\mid t' +\tau - t'' \mid} \phi_j^{\mid t' - t'' \mid}\Big{)}.
\label{eq:computaton_covariance} 
\end{split}
\end{equation}
\end{linenomath}
\end{widetext}

The computation of Equations \ref{eq:computaton_variance} and \ref{eq:computaton_covariance} requires to compute the following three terms: $\sum_{t',t'' = 1}^{T} \phi_k^{\mid t' - t'' \mid} \phi_j^{\mid t' - t'' \mid}$, $\sum_{t',t'' = 1}^{T} \phi_k^{\mid t' +\tau - t'' \mid} \phi_j^{\mid t' -\tau - t'' \mid}$ and $\sum_{t',t'' = 1}^{T} \phi_k^{\mid t' +\tau - t'' \mid} \phi_j^{\mid t' - t'' \mid}$. To solve such terms we point out that a summation of type $\sum_{t',t'' = 1}^{T} (\phi_k \phi_j)^{\mid t' - t'' \mid}$ will result in $T$ points with value $(\phi_k \phi_j)^{0}$, $2(T-1)$ points with value $(\phi_k \phi_j)^{1}$ up to $2(T-t)$ points with value $(\phi_k \phi_j)^{t}$. The summation can be then rewritten as: $\sum_{t',t'' = 1}^{T} (\phi_k \phi_j)^{\mid t' - t'' \mid} = T + \sum_{t=1}^{T-1} (\phi_k \phi_j)^{t} 2 (T-t)$. Similar reasoning can be applied for all the terms above.

\subsubsection{Computation of each summation}
\begin{linenomath}
\begin{equation}
\begin{split}
\texttt{Sum(I)}&: \sum_{t',t'' = 1}^{T} \phi_k^{\mid t' - t'' \mid} \phi_j^{\mid t' - t'' \mid} = T + \sum_{t=1}^{T-1} (\phi_k \phi_j)^{t} 2 (T-t)\\ 
 &= \frac{T-T(\phi_k \phi_j)^{2}+2(\phi_k \phi_j)(\phi_k^{T} \phi_j^{T} - 1)}{(-1 + \phi_k \phi_j)^{2}}.
\end{split}
\label{eq:new_summation_1} 
\end{equation}
\end{linenomath}
\begin{linenomath}
\begin{equation}
\begin{split}
\texttt{Sum}\texttt{(II)}&: \sum_{t',t'' = 1}^{T} \phi_k^{\mid t' + \tau - t'' \mid} \phi_j^{\mid t' - \tau - t'' \mid}\\ 
&= \sum_{t = 1-T}^{T-1} \phi_k^{\mid t+\tau \mid} \phi_j^{\mid t-\tau \mid} (T - \mid t \mid) \\
&= \underbrace{\sum_{t = 1}^{T-1} \phi_k^{(t+\tau)} \phi_j^{\mid t-\tau \mid} (T - t)}_\texttt{Sum(a)}\\ 
&+ \underbrace{\sum_{t = 1 - T}^{0} \phi_k^{\mid t+\tau \mid} \phi_j^{(- t +\tau)} (T + t)}_\texttt{Sum(b)}
\end{split}
\label{eq:new_summation_2} 
\end{equation}
\end{linenomath}
Both summation \texttt{Sum(a)} and \texttt{Sum(b)} can be further split in sums of simple geometric series:
\begin{linenomath}
\begin{equation}
\begin{split}
\texttt{Sum(a)}&: \sum_{t = 1}^{T-1} \phi_k^{(t+\tau)} \phi_j^{\mid t-\tau \mid} (T - t) \\ 
&= \phi_{k}^{\tau}\phi_{j}^{\tau} T \sum_{t=1}^{\tau}(\phi_{k}\phi_{j}^{-1})^t - \phi_{k}^{\tau}\phi_{j}^{\tau} \sum_{t=1}^{\tau} (\phi_{k}\phi_{j}^{-1})^t \cdot t\\
&+ T \phi_{k}^{\tau}\phi_{j}^{-\tau} \sum_{t = \tau + 1}^{T-1} (\phi_{k}\phi_{j})^t - \phi_{k}^{\tau}\phi_{j}^{-\tau} \sum_{t = \tau + 1}^{T-1} (\phi_{k}\phi_{j})^t \cdot t.
\end{split}
\label{eq:new_summation_2_sum_a} 
\end{equation}
\end{linenomath}
\begin{linenomath}
\begin{equation}
\begin{split}
\texttt{Sum(b)}&: \sum_{t = 1 - T}^{0} \phi_k^{\mid t+\tau \mid} \phi_j^{(- t +\tau)} (T + t) \\ 
&= T\phi_{k}^{-\tau}\phi_{j}^{\tau} \sum_{t = 1-T}^{-\tau} (\phi_{k}^{-1}\phi_{j}^{-1})^t\\ 
&+ \phi_{k}^{-\tau}\phi_{j}^{\tau} \sum_{t = 1-T}^{-\tau} (\phi_{k}^{-1}\phi_{j}^{-1})^t \cdot t\\
+& T \phi_{k}^{\tau}\phi_{j}^{\tau} \sum_{t = -\tau + 1}^{0} (\phi_{k}\phi_{j}^{-1})^t \\ 
&+ \phi_{k}^{\tau}\phi_{j}^{\tau}\sum_{t = -\tau + 1}^{0} (\phi_{k}\phi_{j}^{-1})^t \cdot t.
\end{split}
\label{eq:new_summation_2_sum_b} 
\end{equation}
\end{linenomath}
\texttt{Sum(a)} and \texttt{Sum(b)} are composed by geometric series and can be easily solved.\\
\begin{linenomath}
\begin{equation}
\begin{split}
\texttt{Sum}\texttt{(III)}&: \sum_{t',t'' = 1}^{T} \phi_k^{\mid t' + \tau - t'' \mid} \phi_j^{\mid t' - t'' \mid}\\ 
&= \sum_{t = 1-T}^{T-1}\phi_k^{\mid t + \tau \mid}\phi_j^{\mid t \mid} (T - \mid t \mid)\\
&= \underbrace{\sum_{t = 1}^{T-1} \phi_k^{t+\tau}\phi_j^{t}(T-t)}_\texttt{Sum(c)}\\ 
&+ \underbrace{\sum_{t = 1-T}^{0} \phi_k^{\mid t+\tau \mid}\phi_j^{-t}(T+t)}_\texttt{Sum(d)}
\end{split}
\label{eq:new_summation_3} 
\end{equation}
\end{linenomath}
\texttt{Sum(c)} and \texttt{Sum(d)} are composed by geometric series and can be easily solved.\\
\begin{linenomath}
\begin{equation}
\begin{split}
\texttt{Sum}\texttt{(c)}&: \sum_{t = 1}^{T-1} \phi_k^{t+\tau}\phi_j^{t}(T-t)\\ 
&= T \phi_{k}^{\tau}\sum_{t=1}^{T-1} (\phi_{k} \phi_{j})^{t} - \phi_{k} \sum_{t=1}^{T-1} (\phi_{k} \phi_{j})^t \cdot t.
\end{split}
\label{eq:new_summation_3_sum_c} 
\end{equation}
\end{linenomath}
\begin{linenomath}
\begin{equation}
\begin{split}
\texttt{Sum}\texttt{(d)}&: \sum_{t = 1-T}^{0} \phi_k^{\mid t+\tau \mid}\phi_j^{-t}(T+t)\\ 
&= T \phi_k^{-\tau} \sum_{t = 1-T}^{-\tau} (\phi_{k}^{-1}\phi_{j}^{-1})^t + \phi_k^{-\tau} \sum_{t = 1-T}^{-\tau} (\phi_{k}^{-1}\phi_{j}^{-1})^t \cdot t \\
&+ T \phi_k^{\tau} \sum_{t = -\tau + 1}^{0} (\phi_{k}\phi_{j}^{-1})^t + \phi_k^{\tau} \sum_{t = -\tau + 1}^{0} (\phi_{k}\phi_{j}^{-1})^t \cdot t.
\end{split}
\label{eq:new_summation_3_sum_d} 
\end{equation}
\end{linenomath}
\texttt{Sum(c)} and \texttt{Sum(d)} are composed by geometric series and can be easily solved.

\subsubsection{Final result}

We aim in computing the variance of the response estimator $\mathbb{V}\textrm{ar}[R_{k,j}(\tau)]$ as shown in Eq. \ref{eq:pointwise_estimator_R_Variance}. We rewrite the expression in function of the three summations \texttt{Sum}\texttt{(I)}, \texttt{Sum}\texttt{(II)} and \texttt{Sum}\texttt{(III)} solved in the previous section.
\begin{linenomath}
\begin{equation}
\begin{split}
\mathbb{V}\textrm{ar}[R_{k,j}(\tau)] &= \frac{1}{T^2}\Big{(} \texttt{Sum(I)}\\ 
&+ \phi_k^{2\tau} \cdot \texttt{Sum(I)}(\tau = 0)\\ 
&-2 \phi_k^{\tau} \cdot \texttt{Sum(III)} \Big{)}\\ 
&+ \frac{\delta_{k,j}}{T^2} \Big{(}\texttt{Sum(II)}\\ 
&+ \phi_k^{2\tau} \texttt{Sum(II)}(\tau = 0)\\
&- 2 \phi_k^{\tau} \cdot \texttt{Sum(III)}\Big{)}.
\end{split}
\label{eq:pointwise_estimator_variance_summations} 
\end{equation}
\end{linenomath}
Where $\texttt{Sum(I)}(\tau = 0)$ and $\texttt{Sum(II)}(\tau = 0)$ evaluate $\texttt{Sum(I)}$ and $\texttt{Sum(II)}$ in $\tau = 0$.\\

We focus on the asymptotic case $T \gg 1$ and remind the reader that $| \phi_k \phi_j | < 1$. The leading order of the solution is as follows:
\begin{linenomath}
\begin{equation}
\mathbb{V}\textrm{ar}[R_{k,j}(\tau)] = \frac{\phi_k^{2\tau}-1}{T} +\frac{2}{T} \Big{(}\frac{1-\phi_k^{\tau}\phi_j^{\tau}}{1-\phi_k\phi_j} \Big{)} - \frac{2\phi_{k}^{\tau}}{T} \Big{(} \phi_{k} \frac{\phi_{j}^{\tau}-\phi_{k}^{\tau}}{\phi_{j}-\phi_{k}} \Big{)}.
\label{eq:Variance_formula} 
\end{equation}
\end{linenomath}
Finally, we note that in the case of $\phi_{k} = \phi_{j}$ in Eq. \ref{eq:Variance_formula} we substitute the term $\phi_{k}\frac{\phi_{j}^{\tau}-\phi_{k}^{\tau}}{\phi_{j}-\phi_{k}}$ with the limit:
\begin{linenomath}
\begin{equation}
\lim_{\phi_{j}\to\phi_{k}} \phi_{k} \frac{\phi_{j}^{\tau} - \phi_{k}^{\tau}}{\phi_{j} - \phi_{k}} = \phi_{k}^{\tau} \tau.
\label{eq:limit} 
\end{equation}
\end{linenomath}
Equation \ref{eq:Variance_formula} assumes that each time series has been previously normalized to zero mean and unit variance. In case of non-standardized time series, Eq. \ref{eq:Variance_formula} becomes $(\sigma^{2}_k/\sigma^{2}_j) \times \text{Eq. \ref{eq:Variance_formula}}$; $\sigma^{2}_i$ being the variance of time series $x_{i}(t)$ (see also Eq. 15 in \cite{Baldovin}).\\

\section{Confidence bounds. Numerical vs analytical} \label{sec:conf_bounds_test}

\begin{figure*}[tbhp]
\centering
\includegraphics[width=1\linewidth]{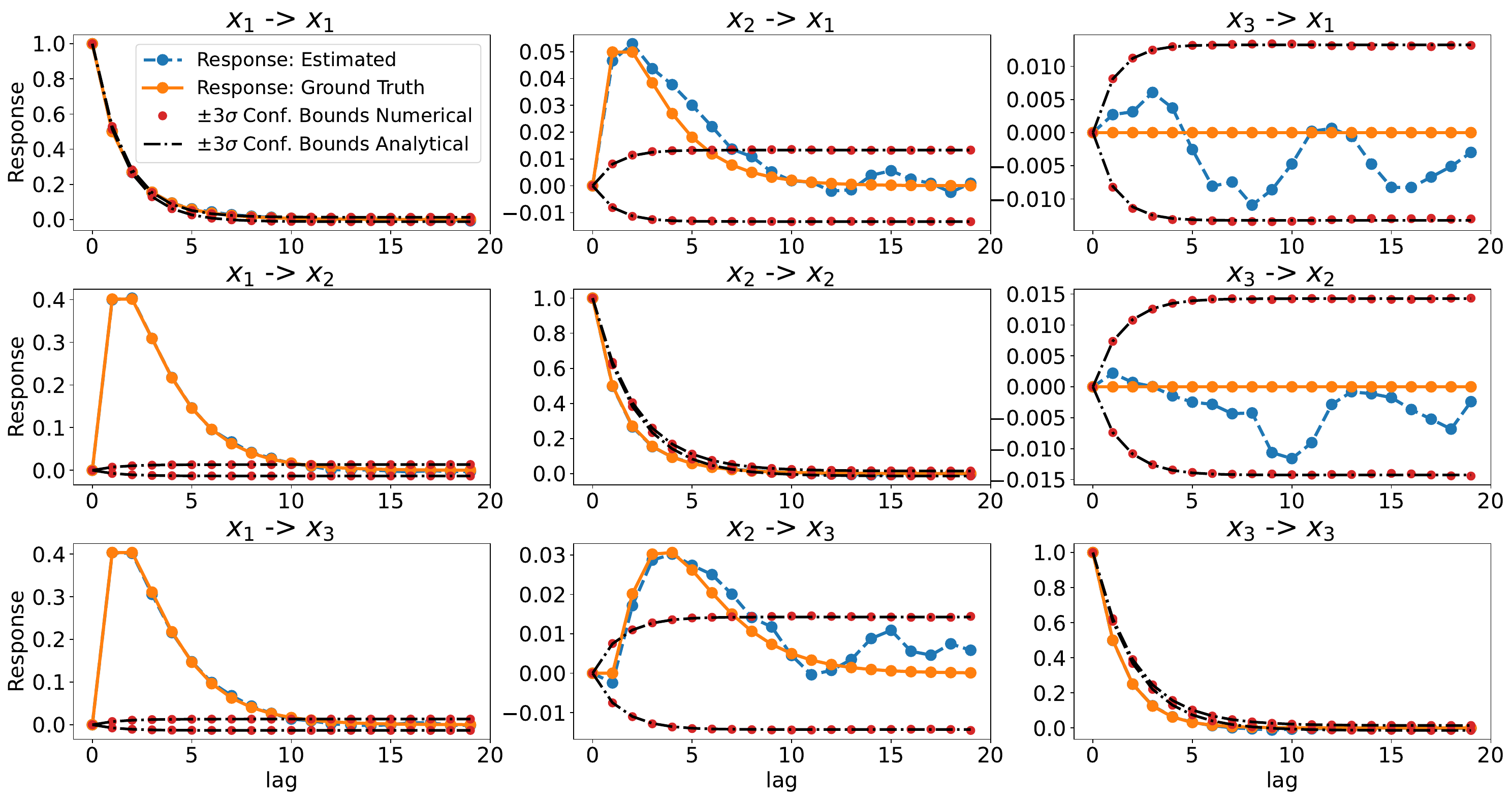}
\caption{Comparing the confidence bounds estimated numerically as in section \ref{sec:numerics_null_model} and the analytical solution as shown in Eq. \ref{eq:expectation_and_Variance_R} for the simple linear Markov model shown in Eq. \ref{eq:linear_markov_matrix_example}. Each panel shows the response $R_{k,j}$ representative of the  causal link $x_j \rightarrow x_k$. All time series have been rescaled to zero
mean and unit variance before computing responses. The response ground truths are shown as solid orange lines. Dashed blue lines are responses estimated through temporal averages: for this step we use a long trajectory of length $T = 10^5$ simulated by system in Eq. \ref{eq:linear_markov_matrix_example}. Red dots indicate the confidence bounds computed numerically using $B = 10^4$ ensemble members of the \textit{null} model as shown in \ref{sec:numerics_null_model}, see section \ref{sec:numerics_null_model}. In each panel, the dot-dashed black line is the analytical solution as in Eq. \ref{eq:expectation_and_Variance_R}. Confidence bounds are set  to $\pm 3 \sigma$. All estimated responses (i.e. blue curves) in between the confidence bounds are here considered as spurious.}
\label{fig:test_temporal_average_all_variables}
\end{figure*}

We consider the system in Eq. \ref{eq:linear_markov_matrix_example} and show all the estimated responses $R_{k,j}$, their ground truths and the confidence bounds in Figure \ref{fig:test_temporal_average_all_variables}. Importantly, we compare the analytical confidence bounds presented in Eq. \ref{eq:expectation_and_Variance_R} with their numerical estimation as shown in section \ref{sec:numerics_null_model}. All bounds are set to $\pm 3 \sigma$.

\clearpage

\section{Histograms of each mode $x_i(t)$ in the global SST field} \label{app:histograms_global}

\begin{figure*}[tbhp]
\centering
\includegraphics[width=0.8\linewidth]{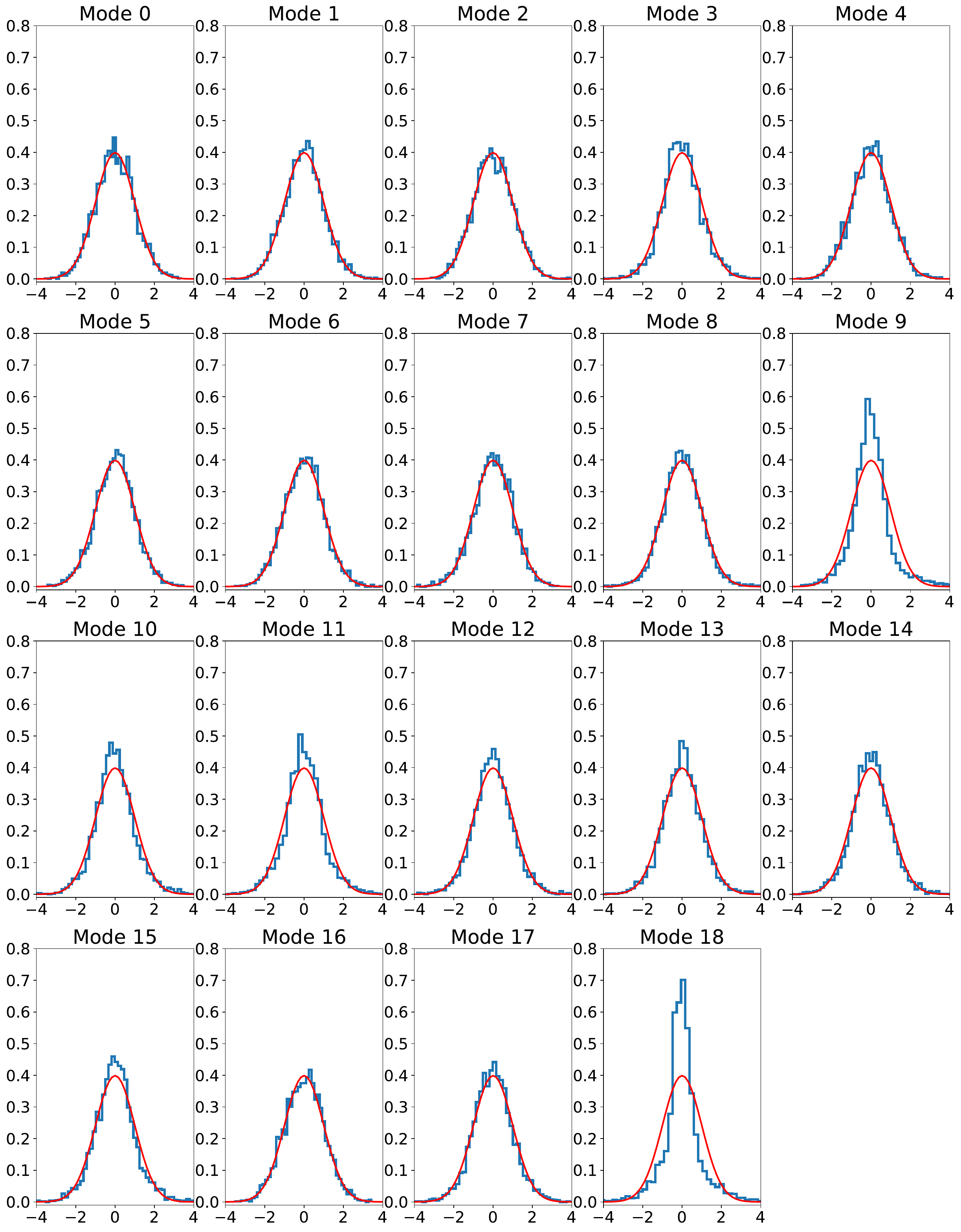}
\caption{Probability distributions of each sea surface temperature signal $x_i(t)$ for each pattern $i$ shown in Fig. \ref{fig:global_communities_strengths}(b). Each signal $x_i(t)$ is first centered to zero mean and standardized to unit variance; therefore the x-axis represents degC per standard deviation. Each pattern (i.e., mode) is here referred to as ``Mode i''. A Gaussian fit is shown in red on top of each histogram.}
\label{fig:histograms_Global}
\end{figure*}

In Fig. \ref{fig:histograms_Global}, we show the histogram of each signal $X(i,t)$ correspondent to pattern $i$ in Fig. \ref{fig:global_communities_strengths}(b). Each $X(i,t)$ has been computed as in Eq. \ref{eq:signals} and it has been then centered to zero mean and standardized to unit variance. A Gaussian distribution with same mean and variance of each $X(i,t)$ is shown in red. The plot shows that the quasi-Gaussian approximation shown in \ref{sec:quasiGaussian} is indeed relevant for the system studied.

\clearpage


\bibliography{apssamp}

\end{document}